\documentclass[lettersize,journal]{IEEEtran}
\usepackage{cite}
\usepackage{amsmath,amssymb,amsfonts}
\usepackage{algorithmic}
\usepackage{graphicx}
\usepackage{textcomp}
\usepackage{xcolor}
\usepackage{mathtools}
\usepackage{ntheorem}
\usepackage{epstopdf}
\usepackage{setspace}
\usepackage{subcaption}
\usepackage{dsfont}
\usepackage{mathrsfs}
\newcommand*{\rom}[1]{\expandafter\@slowromancap\romannumeral #1@}
\usepackage{romannum}
\makeatletter
\renewcommand*\env@matrix[1][*\c@MaxMatrixCols c]{%
\hskip -\arraycolsep
  \let\@ifnextchar\new@ifnextchar
  \array{#1}}
\makeatother
\usepackage{adjustbox}

\newcommand{\Real}{{\mathds R}}

\newcommand{\Nat}{{\mathds N}}

\usepackage{tabularx,booktabs}
\newcolumntype{C}{>{\centering\arraybackslash}X}
\usepackage{lipsum} 

\newtheorem{definition}{Definition}{}
\newtheorem{corollary}{Corollary}{}
\newtheorem{proposition}{Proposition}{}
\newtheorem{problem}{Problem}{}
\newtheorem{theorem}{Theorem}{}
\newtheorem{remark}{Remark}{}
{}
{}
 \usepackage[hidelinks]{hyperref}
 \usepackage{xcolor}
\hypersetup{
    colorlinks,
    linkcolor={blue!80!black},
    citecolor={blue!80!black},
    urlcolor={blue!10!black}
}
\usepackage[noabbrev]{cleveref} 
\creflabelformat{equation}{#2\textup{(\textcolor{red}{#1})}#3}
\creflabelformat{figure}{#2(\textcolor{blue}{#1})#3}

\begin{document}

\title{Privacy in Cloud Computing \\ through Immersion-based Coding}

\author{Haleh Hayati, Nathan van de Wouw, Carlos Murguia%
\thanks{The research leading to these results has received funding from the European Union’s Horizon Europe programme under grant agreement No 101069748 – SELFY project.

Haleh Hayati and Nathan van de Wouw are with the Department of Mechanical Engineering, Dynamics and Control Group, Eindhoven University of Technology, The Netherlands. (Emails: \href{mailto: h.hayati@tue.nl}{h.hayati@tue.nl}; \href{mailto: n.v.d.wouw@tue.nl}{n.v.d.wouw@tue.nl}.)

Carlos Murguia is with the Department of Mechanical Engineering, Dynamics and Control Group, Eindhoven University of Technology, The Netherlands, and with the School of Electrical Engineering and Robotics, Queensland University of Technology, Brisbane, Australia. (Emails: \href{mailto: c.g.murguia@tue.nl}{c.g.murguia@tue.nl}; \href{mailto: carlos.murguia@qut.edu.au}{carlos.murguia@qut.edu.au}.)
}}

\markboth{Journal of \LaTeX\ Class Files,~Vol.~14, No.~8, August~2021}%
{Shell \MakeLowercase{\textit{et al.}}: A Sample Article Using IEEEtran.cls for IEEE Journals}


\maketitle

\begin{abstract}


Cloud computing enables users to process and store data remotely on high-performance computers and servers by sharing data over the Internet. However, transferring data to clouds causes unavoidable privacy concerns. Here, we present a synthesis framework for designing coding mechanisms that allow sharing and processing data in a privacy-preserving manner without sacrificing data utility and algorithmic performance. We consider the setup in which the user aims to run an algorithm in the cloud using private data. The cloud then returns some data utility back to the user (utility refers to the service that the algorithm provides, e.g., classification, prediction, AI models, etc.). To avoid privacy concerns, the proposed scheme provides tools to co-design: 1) coding mechanisms to distort the original data and guarantee a prescribed differential privacy level; 2) an equivalent-but-different algorithm (referred here to as the target algorithm) that runs on distorted data and produces distorted utility; and 3) a decoding function that extracts the true utility from the distorted one with a negligible error. Then, instead of sharing the original data and algorithm with the cloud, only the distorted data and target algorithm are disclosed, thereby avoiding privacy concerns. The proposed scheme is built on the synergy of differential privacy and system immersion tools from control theory. The key underlying idea is to design a higher-dimensional target algorithm that embeds all trajectories of the original algorithm and works on randomly encoded data to produce randomly encoded utility. We show that the proposed scheme can be designed to offer any level of differential privacy without degrading the algorithm's utility. We present two use cases to illustrate the performance of the developed tools: privacy in optimization/learning algorithms and a nonlinear networked control system. \end{abstract}


%
\IEEEpeerreviewmaketitle

\section{Introduction}
Scientific and technological advancements in today's hyper-connected world have resulted in a massive amount of user data being collected and processed by hundreds of companies over public networks and servers. Companies use these data to provide personalized services and targeted advertising. However, this technology has come at the cost of an alarming loss of privacy in society. Depending on their resources, adversaries can infer sensitive (private) information about users from public data available on the Internet and/or unsecured networks/servers. A motivating example of this privacy loss is the potential use of data from smart electrical meters by criminals, advertising agencies, and governments for monitoring the presence and activities of occupants \cite{Poor1,Poor2}. Other examples are privacy loss caused by information sharing in distributed control systems and cloud computing \cite{Huang:2014:CDP:2566468.2566474}; the use of travel data for traffic estimation in intelligent transportation systems \cite{Gruteser}; and data collection and sharing by the Internet-of-Things (IoT) \cite{WEBER201023}. These, and many more, privacy challenges have drawn the interests of researchers from different fields (e.g., information theory, computer science, and control engineering) to the broad research subject of privacy and security in data-sharing and cloud computing, see \cite{xiao2012security,Jerome1,ferrari2021safety,Takashi_3,Carlos_Iman1,wang2024secure,prakash2024privacy}.\\
\indent In recent years, various privacy-preserving schemes have been proposed to address privacy concerns in data-sharing and cloud computing. Most of them rely on perturbation techniques (inject randomness to distort data) or cryptographic tools (encrypt data/algorithms before disclosure), see \cite{farokhi2020privacy}.\\
\indent Perturbation-based techniques use information-theoretic privacy metrics \cite{Farokhi1,murguia2021privacy,hayati2021finite,10384020} and/or differential privacy \cite{Jerome1,Dwork}. In these techniques, random vectors drawn from known probability distributions are injected into sensitive data before sharing it with the cloud. Although these methods effectively reduce the amount of private data that can be inferred, the induced distortion is never removed, which inevitably leads to performance degradation of the utility when used to run algorithms in the cloud.\\
\indent In standard cryptographic methods, data is encrypted before sharing and then decrypted in the cloud before processing, see \cite{wan2018physical}. This technique is suitable for making eavesdropping attacks over communication networks difficult. However, it is still vulnerable to insider attacks in the cloud (as they have the decryption key). Alternatively, Homomorphic Encryption (HE) methods do not require data to be decrypted before processing. They allow computations over plain data by performing appropriate computations on the encrypted data \cite{Gatsis_homo,joshi2022comparative}. However, standard HE methods (e.g., Paillier and ElGamal encryption \cite{paillier1999public}) work over finite rings of integers, which implies that the original algorithm must be reformulated to also work on rings of integers \cite{murguia2020secure}. This reformulation, although effective when possible, is hard to accomplish for general algorithms. In the literature, it has only been successfully performed on small classes of algorithms (mainly algorithms with linear transitions and a fairly small number of variables). Even when the reformulation is possible, mapping algorithms working on the reals to operate on finite rings leads to performance degradation and large computational overhead \cite{kim2022comparison}.\\
\indent In summary, although current solutions improve privacy of data and algorithms, they often do it at the expense of data utility and algorithm performance degradation. Balancing these tradeoffs is a key challenge when implementing privacy solutions for cloud computing. Hence, novel privacy-preserving schemes must be designed to provide strict privacy guarantees with a fair computational cost and without compromising the application performance and utility.\\
\indent The aim of this work is to devise synthesis tools that allow designing coding mechanisms that protect private information and allow the implementation of remote \emph{dynamic algorithms} -- algorithms that perform iterations and have memory. We propose a coding scheme built on the synergy of random coding and \emph{system immersion} tools \cite{astolfi2003immersion} from control theory. The key idea is to treat the original algorithm as a dynamical system that we seek to \emph{immerse} into a higher-dimensional system (the so-called target algorithm), i.e., trajectories of the target algorithm must embed all trajectories of the original (up to a random (left-)invertible transformation). The target algorithm must be designed such that: 1) trajectories of the original algorithm are immersed/embedded in its trajectories, and 2) it operates on randomly encoded higher-dimensional data to produce randomly encoded utility. We formulate the coding mechanism at the user side as a random change of coordinates that maps original private data to a higher-dimensional space. Such coding enforces that the target algorithm produces an encoded higher-dimensional version of the utility of the original algorithm. The encoded utility is decoded at the user side using a left-inverse of the encoding map. This is basically a \emph{homomorphic encryption scheme that operates over the reals} instead of finite rings of integers (as in the standard case). Working over the reals provides much more freedom to redesign algorithms to work on encoded/encrypted data, which translates to our scheme being able to tackle a larger class of (nonlinear and high-dimensional) dynamic algorithms.\\
\indent Another difference compared to HE schemes is the privacy guarantee. While standard HE schemes provide hard security guarantees, the scheme we propose gives arbitrarily strong probabilistic guarantees (in terms of differential privacy). We show that our scheme provides the same utility as the original algorithm (i.e., when no coding is employed to protect against data inference), (practically) reveals no information about private data, can be applied to large-scale algorithms (hundreds of thousands of algorithm variables, as commonly occurs in deep-learning and large-scale simulations), is computationally efficient, and offers any desired level of differential privacy without degrading the algorithm utility (because the induced distortion is removed at the user side).\\
\indent Summarizing, the main contribution of the paper is a prescriptive synthesis framework (i.e., it uses the original undistorted algorithm as the basis) that allows the joint design of coding schemes to randomize data before disclosure and new (target) algorithms that work on encoded data, does not lead to performance degradation, and provides any desired level of differential privacy. 
{The concept of immersion-based coding scheme was initially introduced in our previous works \cite{hayati2022privacy}, \cite{hayati2023mo}, \cite{hayati2023immersion}. In those studies, we established the foundational concepts and demonstrated their application specifically in "privacy-preserving federated learning" and "privacy-preserving remote anomaly detection". This paper builds on our earlier research by significantly extending and generalizing the immersion-based coding scheme to support various dynamic algorithms. These algorithms are modeled as sets of difference/differential parameter-varying equations that process user data to produce utility. Furthermore, we introduce a novel aspect by integrating arbitrarily strong probabilistic guarantees in terms of differential privacy into the scheme. In particular, we prove that the proposed scheme can provide any desired level of differential privacy without reducing the accuracy and performance of the original algorithm, which is a key contribution distinguishing our current work from our preliminary studies. Furthermore, we demonstrate the performance of the proposed tools for two use cases: privacy in optimization/learning algorithms and nonlinear networked control systems.}\\
\indent The remainder of this paper is organized as follows. In Section \ref{sec2}, we formulate the problem of designing the immersion-based coding for privacy of dynamical algorithms. In Section \ref{sec3}, we construct a prescriptive solution design to the proposed problem using random affine maps. In Section \ref{sec4}, we prove that the proposed solution can provide any desired level of differential privacy. In Section \ref{sec5}, we provide a synthesis procedure to summarize the operation methodology for the proposed mechanism. In Section \ref{sec6}, the performance of the immersion-based coding is shown for two use cases of privacy in optimization/learning algorithms and nonlinear networked control systems. Conclusions are drawn in Section \ref{sec7}.

\emph{Notation}: The symbol $\Real$ represents the real numbers and $\Real^+$ denotes the set of positive real numbers. The symbol $\Nat$ represents the set of natural numbers and $\Nat_0$ denotes the set of natural numbers with zero. For $x \in \Real^n$, the Euclidian norm is denoted by $||x||$, $||x||^2=x^{\top}x$, where $^{\top}$ denotes transposition, and the $l^1$ norm as $\|x\|_1 = \sum_{i=1}^{N}|x^i|$, $x = (x^1,\ldots,x^N)^\top$. The $n \times n$ identity matrix is denoted by $I_n$ or simply $I$ if $n$ is clear from the context. Similarly, $n \times m$ matrices composed of only ones and only zeros are denoted by $\mathbf{1}_{n \times m}$ and $\mathbf{0}_{n \times m}$, respectively, or simply $\mathbf{1}$ and $\mathbf{0}$ when their dimensions are clear. 
The notation $x\sim \mathcal{N}[\mu^x,\Sigma^x]$ means $x \in \Real^{n}$ is a normally distributed random vector with mean $E[x] = \mu^x \in \Real^{n}$ and covariance matrix $E[(x-\mu^x)(x-\mu^x)^\top] = \Sigma^x \in \Real^{n \times n}$, where $E[a]$ denotes the expected value of the random vector $a$. The notation $y \sim \operatorname{Laplace }(\mu^y, \Sigma^y )$ stands for the Laplace random vector $ y\in \Real^{m}$ with mean $E(y)=\mu^y$ and scale parameter $\frac{1}{2}E[(y-\mu^y)(y-\mu^y)^\top] = \Sigma^y \in \Real^{m \times m}$. The left inverse of matrix $X$ is denoted by $X^L$ where $X^L X = I$.
\section{Problem Formulation}\label{sec2}
The setting we consider is the following. The user aims to run an algorithm remotely (in the cloud) using her/his local data $y_k$ to produce some data utility $u_k$ (what the algorithm returns); see Fig. \ref{flowchart}(a). However, the user does not want to share the original data $y_k$, as they contain sensitive private information that could be inferred. The goal of our proposed privacy-preserving scheme is to devise an encoding mechanism to mask the original data \emph{and the algorithm to be run in the cloud} so that inference of sensitive information from the encoded disclosed data is as hard as possible (in a probabilistic Differential-Privacy (DP) sense) \emph{without degrading the performance of the encoded algorithm} (so a homomorphic probabilistic scheme).\\
\indent As the original algorithm running in the cloud, we consider discrete-time dynamic algorithms of the form:
\begin{equation}\label{generaldynamics}
\Sigma=\left\{\begin{array}{l}
\begin{aligned}
\zeta_{k+1}&=f\left(\zeta_k, y_k,w_k\right), \\
u_k&=g\left(\zeta_k, y_k,w_k \right),
\end{aligned}
\end{array}\right.
\end{equation}
with time-index $k \in \Nat_0$, algorithm input data $y_k \in \mathbb{R}^{n_y}$ (the user data), generated data utility $u_k \in \mathbb{R}^{n_u}$, algorithm internal variables $\zeta_k \in \mathbb{R}^{n_\zeta}$, unknown disturbance (or reference signal) $w_k\in \mathbb{R}^{n_w}$, and functions $f:\mathbb{R}^{n_\zeta} \times \mathbb{R}^{n_y} \times \mathbb{R}^{n_w} \rightarrow \mathbb{R}^{n_\zeta}$ and $g:\mathbb{R}^{n_\zeta} \times \mathbb{R}^{n_y}  \times \mathbb{R}^{n_w} \rightarrow \mathbb{R}^{n_u}$. This algorithm description covers a large class of algorithms that have memory and recursively converge to some value of interest -- e.g., algorithms for learning, searching, simulation, control, and estimation.\\
\indent To prevent adversaries from eavesdropping on the network or in the cloud itself to access private data, we propose a privacy-preserving coding scheme to distort data before transmission. Let the user distort the original data before disclosure through some encoding map $\pi_1:\mathbb{R}^{n_y} \rightarrow \mathbb{R}^{{\tilde{n}_y}}$, $\tilde{y}_k=\pi_1(y_k)$, and share the distorted $\tilde{y}_k$ with the cloud to run the algorithm. In general, running the original algorithm $\Sigma$ on the distorted data $\tilde{y}_k$ will not yield the same utility $u_k$ that would be obtained if it was run using $y_k$. That is, privacy-preserving methods that do not account for (remove) the distortion induced by the encoding map $\pi_1(\cdot)$ lead to performance degradation. This is precisely the problem with perturbation-based techniques for privacy preservation (like with standard, non-homomorphic DP tools).\\
\indent To tackle these challenges, the scheme, proposed here, seeks to design a new dynamic algorithm $\Tilde{\Sigma}$ (referred hereafter to as the target algorithm) that runs on encoded data $\tilde{y}_k$, operates over the reals, and returns an encoded utility, $\Tilde{u}_k$, that can be used to extract the true utility $u_k$ at the user side without distortion, i.e., a homomorphic encryption scheme that operates over the reals. Letting the encoding map and target algorithm work over the reals (as opposed to standard HE schemes that operate over finite rings of integers) provides much more freedom to design $\pi_1(\cdot)$ and $\Tilde{\Sigma}$, which will allow tackling a larger class of (nonlinear and high-dimensional) dynamic algorithms.\\
\indent We seek to design $\pi_1(\cdot)$ such that $\tilde{y}_k = \pi_1(y_k) \in \mathbb{R}^{\tilde{n}_y}$ is of higher dimension than $y_k \in \mathbb{R}^{n_y}$, i.e., $\tilde{n}_y > n_y$. We impose this condition to create redundancy in both the encoding map and target algorithm. Redundancy will allow us to inject randomness that can be traced through the algorithm, removed at the user side, and used to enforce an arbitrary level of differential privacy. In addition, embedding the original dynamical algorithm into a higher-dimensional target algorithm can effectively protect the system's dimension, which includes crucial and sensitive private information. Consider the higher-dimensional target algorithm:
\begin{equation}\label{targetalgorithm}
 \tilde{\Sigma}=\left\{\begin{array}{l}\begin{aligned}
\tilde{\zeta}_{k+1}&=\tilde{f}(\tilde{\zeta}_k, \tilde{y}_k,w_k), \\
\tilde{u}_k&=\tilde{g}(\tilde{\zeta}_k, \tilde{y}_k,w_k),
\end{aligned}
\end{array}\right.
\end{equation}
with distorted user data $\tilde{y}_k \in \mathbb{R}^{\tilde{n}_y}$, $\tilde{n}_y > n_y$, distorted data utility $\tilde{u}_k \in \mathbb{R}^{\tilde{n}_u}$, $\tilde{n}_u > n_u$, generated by the algorithm, algorithm internal variables $\tilde{\zeta}_k \in \mathbb{R}^{\tilde{n}_\zeta}$, $\tilde{n}_\zeta > n_\zeta$, and functions $\tilde{f}:\mathbb{R}^{\tilde{n}_\zeta} \times \mathbb{R}^{\tilde{n}_y} \times \mathbb{R}^{n_w} \rightarrow \mathbb{R}^{\tilde{n}_\zeta}$ and $\tilde{g}:\mathbb{R}^{\tilde{n}_\zeta} \times \mathbb{R}^{\tilde{n}_y} \times \mathbb{R}^{n_w} \rightarrow \mathbb{R}^{\tilde{n}_u}$.
Our goal is to design the encoding map $\pi_1(\cdot)$ and the functions $\Tilde{f}(\cdot)$ and $\Tilde{g}(\cdot)$ such that $\Tilde{\Sigma}$ can work on the encoded data $\Tilde{y}_k=\pi_1(y_k)$ to produce encoded utility $\Tilde{u}_k$ that can be used to extract $u_k$.\\
\indent At a system-theoretic level, what we seek to accomplish is to embed trajectories $(\zeta_k,u_k)$ of the original algorithm $\Sigma$ (in response to the user data $y_k$) into trajectories $(\tilde{\zeta}_k,\tilde{u}_k)$ of the target algorithm (in response to the encoded user data $\tilde{y}_k$). That is, we aim to design $\pi_1(\cdot)$ and $\Tilde{\Sigma}$ so that there exists a bijection between trajectories of both algorithms (referred here to as the \emph{immersion map}), and thus, having a trajectory of the target algorithm uniquely determines a trajectory of the original one through the immersion map. The latter hints at the possibility of running the target algorithm in the cloud (instead of the original) on encoded data $\tilde{y}_k$, and using its trajectories $(\tilde{\zeta}_k,\tilde{u}_k)$ to extract ${u}_k$. So a fundamental question is how do we design $\pi_1(\cdot)$ and $\Tilde{\Sigma}$ to accomplish this bijection?\\
\indent In system and control theory, this type of embedding between systems trajectories is referred to as \emph{system immersion} and has been used for nonlinear adaptive control \cite{astolfi2003immersion,AstolfiBook} and output regulation \cite{IsidoriRegulation,IsidoriBook}. However, using system immersion tools in the context of privacy in cloud computing has been largely unexplored in the literature. In what follows, we discuss the necessary mathematical machinery and provide sufficient conditions to simultaneously design the encoding map $\pi_1(\cdot)$ and the target algorithm $\Tilde{\Sigma}$ to accomplish immersion and utility extraction using ideas from system immersion. This will culminate in a problem description on immersion-based coding for privacy at the end of this section.

\subsection{Immersion-based Coding}
Consider the original and target algorithms, $\Sigma$ and $\tilde{\Sigma}$, in \eqref{generaldynamics} and \eqref{targetalgorithm}, respectively. We say that $\Sigma$ is immersed in $\tilde{\Sigma}$ if $\tilde{n}_{\zeta} > {n}_{\zeta}$ and there exists a function $\pi_2: \mathbb{R}^{n_\zeta} \rightarrow\mathbb{R}^{\tilde{n}_\zeta}$ that satisfies $\tilde{\zeta}_k=\pi_2(\zeta_k)$ for all $\zeta_k$ and $\tilde{\zeta}_k$ generated by $\Sigma$ and $\tilde{\Sigma}$, respectively. That is, any trajectory of $\tilde{\Sigma}$ is a trajectory of $\Sigma$ through the mapping $\pi_2(\cdot)$, and $\pi_2(\cdot)$ is an immersion because the dimension of its image is ${\tilde{n}_\zeta}>{{n}_\zeta}$.
We refer to this map $\pi_2(\cdot)$ as the \emph{immersion map}.\\
\begin{figure*}[t]
  \centering
  \subcaptionbox{}[.43\linewidth]{%
    \includegraphics[width=0.9\linewidth]{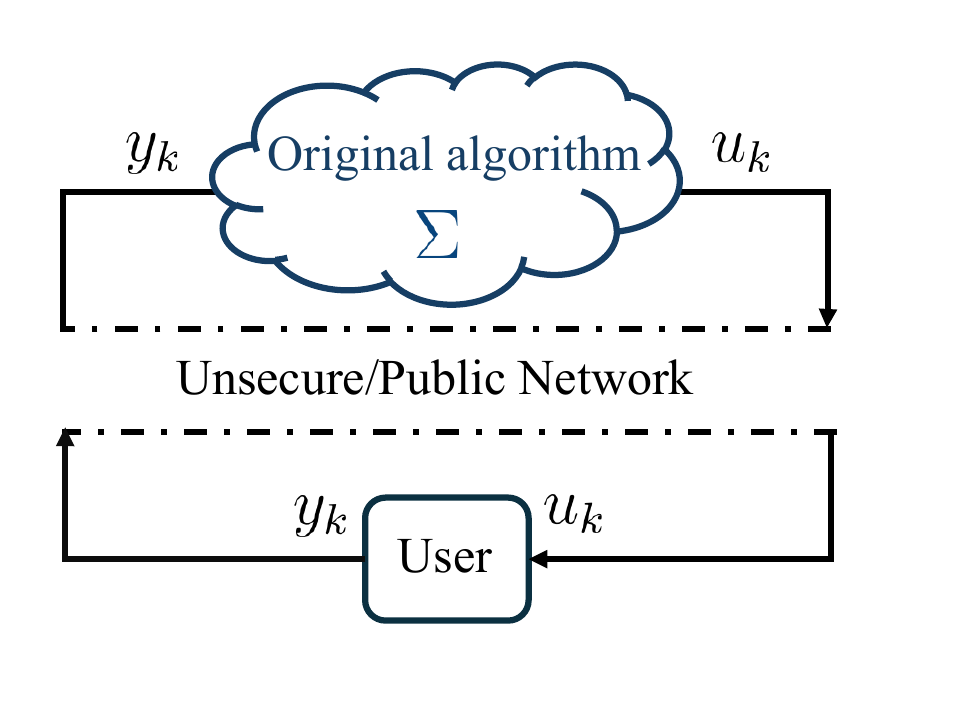}}\quad
  \subcaptionbox{}[.48\linewidth]{%
    \includegraphics[width=0.8\linewidth]{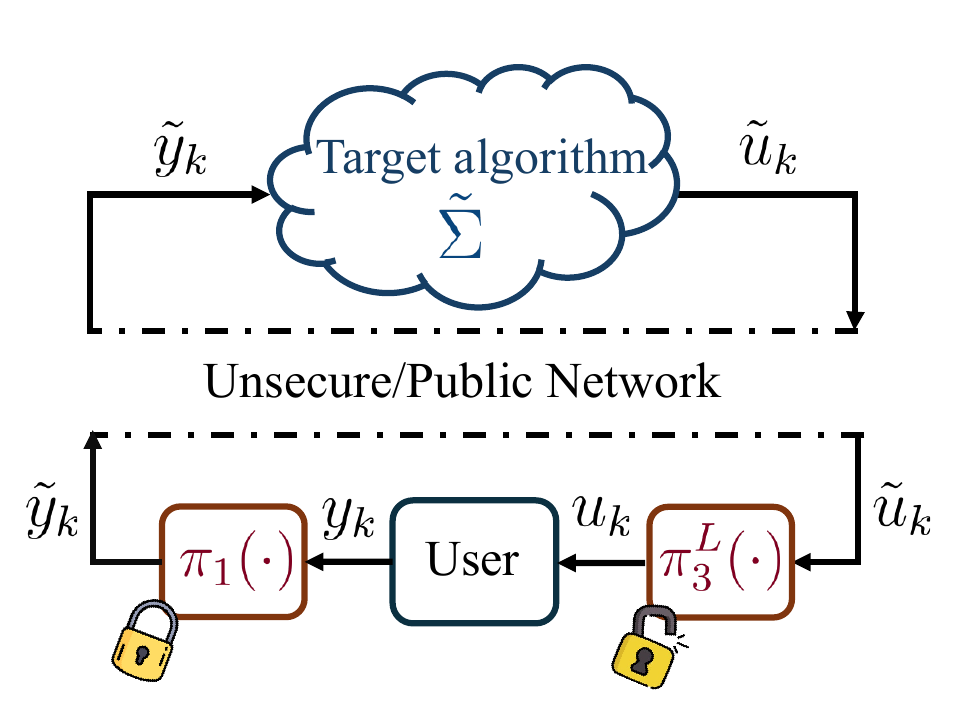}}
    \caption{The schematic diagram of a networked dynamical algorithm (a) without privacy and (b) with immersion-based coding for privacy.}
\label{flowchart}
\end{figure*}
\indent To guarantee that $\Sigma$ is immersed in $\Tilde{\Sigma}$ (in the sense introduced above), we need to impose conditions on the functions shaping the algorithms, their initial conditions, and the encoding and immersion maps, $(f, g, \zeta_{0},\pi_1, \pi_2, \tilde{f},\tilde{g},\tilde{\zeta}_{0})$. In particular, we require designing $(\pi_1, \pi_2, \tilde{f},\tilde{\zeta}_{0})$ such that $\tilde{\zeta}_k=\pi_2(\zeta_k)$ for all $k$ and $(\zeta_k,\tilde{\zeta}_k) \in  \mathbb{R}^{{n}_\zeta}\times \mathbb{R}^{\tilde{n}_\zeta}$, i.e., the manifold $\tilde{\zeta}_k=\pi_2(\zeta_k)$ must be forward invariant under the algorithms in \eqref{generaldynamics} and \eqref{targetalgorithm}. Define the off-the-manifold error $e_k:=\tilde{\zeta}_k-\pi_2(\zeta_k)$. The manifold $\tilde{\zeta}_k=\pi_2(\zeta_k)$ is forward invariant if and only if the origin of the error dynamics:
\begin{equation}
\begin{aligned}\label{errordynamic}
&{e}_{k+1}=\tilde{\zeta}_{k+1}- \pi_2(\zeta_{k+1})\\
&=\tilde{f}(e_k+\pi_2(\zeta_k),\pi_1(y_k),w_k)-\pi_2(f({\zeta}_k,{y_k},w_k)),
\end{aligned}
\end{equation}
is a fixed point, i.e., $e_k=\mathbf{0}$ implies $e_{k+1}=\mathbf{0}$ for all $k \in  \Nat_0$ \cite{hirsch2012differential}. 
Substituting $e_k=\mathbf{0}$ and $e_{k+1}=\mathbf{0}$ in \eqref{errordynamic} leads to
\begin{equation} \label{dynamicsmap}
    \tilde{f}(\pi_2(\zeta_k),\pi_1(y_k),w_k)-\pi_2(f({\zeta}_k,{y}_k,w_k))=\mathbf{0}.
\end{equation}
Therefore, $\tilde{\zeta}_k=\pi_2(\zeta_k)$ is satisfied for all $k$ if: \textbf{(1)} the initial condition of \eqref{targetalgorithm}, $\tilde{\zeta}_0$, satisfies $\tilde{\zeta}_0 = \pi_2({\zeta}_0)$, which leads to $e_0=\mathbf{0}$ (start on the manifold); and \textbf{(2)} the dynamics of both algorithms match under the immersion and encoding maps, i.e., \eqref{dynamicsmap} is satisfied (invariance condition on the manifold). We refer to these two conditions as the \emph{immersion conditions}.
\\[1mm]
\noindent\makebox[\linewidth]{\rule{\linewidth}{0.8pt}}
\textbf{Immersion Conditions:}\
\begin{equation}\label{immersionconditiongeneral}
    \left\{
\begin{aligned}
&\tilde{f}(\pi_2(\zeta_k),\pi_1(y_k),w_k)=\pi_2(f({\zeta}_k,{y}_k,w_k)), \text{ (invariance)}\\
&\tilde{\zeta}_0=\pi_2({\zeta}_0). \text { (start on the manifold)}
\end{aligned}
\right. \vspace{-2.5mm}
\end{equation}
\noindent\makebox[\linewidth]{\rule{\linewidth}{0.8pt}}

So far, we have derived sufficient conditions \eqref{immersionconditiongeneral} for the trajectories of the original algorithm to be immersed in the trajectories of the target algorithm in terms of $(f,\tilde{f},{\zeta}_0,{\tilde{\zeta}}_0,\pi_1,\pi_2)$. Next, we derive conditions on the utility maps $(g,\tilde{g})$ so that, given the distorted utility $\tilde{u}_k$, the true utility $u_k$ can be extracted at the user side. That is, we need conditions under which there exists a \emph{left-invertible mapping} $\pi_3: \mathbb{R}^{n_u} \rightarrow \mathbb{R}^{\tilde{n}_u}$ satisfying $\tilde{u}_k = \pi_3(u_k)$ for all $k$, which leads to $u_k=\pi_3^L(\tilde{u}_k)$ for some left-inverse function $\pi_3^L(\cdot)$ of $\pi_3(\cdot)$, i.e., $\pi_3^L \circ \pi_3 (s) = s$. By substituting the expressions of $u_k$ and $\tilde{u}_k$, in \eqref{generaldynamics}-\eqref{targetalgorithm}, and the coding and immersion maps, $\tilde{y}_k=\pi_1(y_k)$ and $\tilde{\zeta}_k=\pi_2(\zeta_k)$, we have that $\tilde{u}_k = \pi_3(u_k)$ implies
\begin{equation}\label{leftinvertibilitygeneral1}
    \pi_{3}({g}({\zeta}_k, {y}_k,w_k))=\tilde{g}(\pi_2(\zeta_k), \pi_1(y_k),w_k)).
\end{equation}
Therefore, to be able to extract $u_k$ from $\tilde{u}_k$, we need to design $(\tilde{g},\pi_1,\pi_2)$ so that \eqref{leftinvertibilitygeneral1} is satisfied for given $g(\cdot)$, all $\zeta_k$, $y_k$, and $w_k$, and some left-invertible mapping $\pi_3(\cdot)$.

We now have all the ingredients to pose the problem we seek to address.

\begin{problem}\label{problem1}\emph{\textbf{(Immersion-based Coding for Privacy)}} For given $(f,g,\zeta_0)$ of the original algorithm \eqref{generaldynamics}, design an encoding map $\pi_1(\cdot)$, immersion map $\pi_2(\cdot)$, and $(\tilde{f},\tilde{g},\tilde{\zeta}_0)$ of the target algorithm \eqref{targetalgorithm} so that: \textbf{(a)} the immersion conditions \eqref{immersionconditiongeneral} hold; and \textbf{(b)} there exists a left-invertible function $\pi_3(\cdot)$ satisfying \eqref{leftinvertibilitygeneral1}.
\end{problem}

The flowchart of the original algorithm and the target algorithm are shown in Fig. \ref{flowchart}.

\begin{remark}
Note that solutions to Problem \ref{problem1} characterize a class of encoding maps and target algorithms for which we can design homomorphic encryption schemes (a prescriptive design, i.e., for given $(f,g,\zeta_0)$) that operate over the reals. However, this class is infinite-dimensional (over a function space). It leads to an underdetermined algebraic problem with an infinite dimensional solution space. To address this aspect of the problem, we impose structure on the maps we seek to design. We restrict to random affine maps composed of linear coordinate transformations and additive random processes. In what follows, we prove that this class of maps is sufficient to guarantee an arbitrary level of differential privacy.
\end{remark}

\section{Affine Solution to Problem \ref{problem1}}\label{sec3}
In this section, we construct a prescriptive solution to Problem \ref{problem1} using random affine maps. Let the user data encoding map $\pi_1(\cdot)$, the immersion map $\pi_2(\cdot)$, and the utility map $\pi_3(\cdot)$ be affine functions of the form:
\begin{equation}\label{privacymechanism2}
    \left\{\begin{aligned}
    \tilde{y}_k &=\pi_1(y_k) := \Pi_1 y_k+b^1_k, \\[-.5mm]
    \tilde{\zeta}_k &=\pi_2(\zeta_k) := \Pi_2 \zeta_k, \\[-.5mm]
    \tilde{u}_k  &=\pi_3(u_k) := \Pi_3 u_k+b^3_k,
    \end{aligned}\right.
\end{equation}
for matrices $\Pi_1 \in \mathbb{R}^{\tilde{n}_y \times n_y}$, $\Pi_2 \in \mathbb{R}^{\tilde{n}_\zeta \times n_\zeta}$, and $\Pi_3 \in \mathbb{R}^{\tilde{n}_u \times n_u}$ to be designed, with $\tilde{n}_y> n_y$, $\tilde{n}_u> n_u$, $\tilde{n}_\zeta> n_\zeta$, and some i.i.d. multivariate random processes $b^1_k \in \mathbb{R}^{\tilde{n}_y}$ and $b^3_k \in \mathbb{R}^{\tilde{n}_u}$. Therefore, for these maps, the immersion conditions \eqref{immersionconditiongeneral} amounts to $\tilde{\zeta}_0=\Pi_2\zeta_0$ and
\begin{equation}\label{immersioncondition2}
    \begin{aligned}
    &\tilde{f}\left(\Pi_2 \zeta_k,\Pi_1 y_k+b^1_k,w_k\right)=\Pi_{2} f\left({\zeta}_k,{y}_k,w_k\right).
\end{aligned}
\end{equation}
Let the function $\tilde{f}(\cdot)$ be of the form:
\begin{equation}\label{eq:20}
    \tilde{f}\left(\tilde{\zeta}_k, \tilde{y}_k,w_k\right) = M_3 f\left(M_2\tilde{\zeta}_k, M_1\tilde{y}_k,w_k\right),
\end{equation}
for some matrices $M_1 \in \mathbb{R}^{n_y \times \tilde{n}_y}$, $M_2 \in \mathbb{R}^{n_\zeta \times \tilde{n}_\zeta}$, and $M_3 \in \mathbb{R}^{\tilde{n}_\zeta \times n_\zeta }$ to be designed. Hence, the immersion condition \eqref{immersioncondition2} takes the form:
\begin{equation}\label{eqc1}
M_3 f\left(M_2\tilde{\zeta}_k, M_1\tilde{y}_k,w_k\right) = \Pi_{2} f\left({\zeta}_k,{y}_k,w_k\right).
\end{equation}
Note that the choice of $\tilde{f}$ in \eqref{eq:20} provides a prescriptive design in terms of the original algorithm function $f(\cdot)$. That is, we exploit the knowledge of the original algorithm and build the target algorithm on top of it in an algebraic manner.\\
\indent To satisfy \eqref{eqc1}, we must enforce $M_1\left(\Pi_1 y_k +b^1_k\right)=y_k$, $M_2 \Pi_2 {\zeta}_k ={\zeta}_k$, and $M_3=\Pi_{2}$, which implies that $M_1 \Pi_1=I$, $M_2 \Pi_2=I$, and $M_1 b^1_k=\mathbf{0}$ (i.e., $b^1_k \in \text{ker}[M_1]$). From this brief analysis, we can draw the following conclusions: 1) $\Pi_1$ and $\Pi_2$ \emph{must be of full column rank} (i.e., $\text{rank}[\Pi_1]=n_y$ and $\text{rank}[\Pi_2]=n_\zeta$); 2) $M_1$ and $M_2$ are left inverses of $\Pi_1$ and $\Pi_2$, i.e.  $M_1 = \Pi^L_1$ and $M_2 = \Pi^L_2$ (which always exist given the rank of $\Pi_1$ and $\Pi_2$); 3) $b^1_k \in \text{ker}[\Pi^L_1]$ (this kernel is always nontrivial because $\Pi_1$ is full column rank by construction); and 4) $M_3=\Pi_{2}$. Combining all these facts, the final form of $\tilde{f}(\cdot)$ in \eqref{targetalgorithm} is given by
\begin{equation}\label{eqc2}
\tilde{f}\big(\tilde{\zeta}_k, \tilde{y}_k,w_k\big) = \Pi_2 f\big(\Pi^L_2 \tilde{\zeta}_k, \Pi^L_1\tilde{y}_k,w_k\big).
\end{equation}
At every $k$, vector $b^1_k$ is designed to satisfy $\Pi_{1}^{L} b^1_k=\mathbf{0}$ and used to distort the coding map $\pi_1(\cdot)$ in \eqref{privacymechanism2}. The role of $b^1_k$ is to randomize the data so that we can guarantee differential privacy. The idea is that because the user knows that $b^1_k$ draws realizations from the kernel of $\Pi_{1}^{L}$, the distortion induced by it can be removed at the user side (following the same reasoning as standard HE, but operating over the reals and proving a DP guarantee). To enforce that $b^1_k \in \text{ker}[\Pi_{1}^{L}]$, without loss of generality, we let it be of the form $b^1_k=N_1 s_k$
for some matrix $N_1 \in \mathbb{R}^{\tilde{n}_y \times (\tilde{n}_y - n_y)}$ expanding the kernel of $\Pi_{1}^{L}$ (i.e., $\Pi_{1}^{L}N_1=\mathbf{0}$) and an arbitrary i.i.d. process $s_k \in \mathbb{R}^{(\tilde{n}_y - n_y)}$. This structure for $b^1_k$ always satisfies $\Pi_{1}^{L}b^1_k = \Pi_{1}^{L}N_1 s_k = \mathbf{0}$, for $s_k$ with arbitrary probability distribution.\\
\indent Up to this point, we have designed $(\tilde{f},\pi_1,\pi_2)$ to satisfy immersion conditions \eqref{immersionconditiongeneral} for the affine maps in \eqref{privacymechanism2}. Next, we seek to design the utility map $\tilde{g}(\cdot)$ to satisfy the encoding introduced in \eqref{privacymechanism2} ($\tilde{u}_k = \Pi_3 u_k+b^3_k$) and so that there exists a decoding function that extracts the true utility $u_k$ from the encoded $\tilde{u}_k$ (i.e., the left-invertibility of $\pi_3(\cdot)$). To design $\tilde{g}(\cdot)$, similarly to what we did for $\tilde{f}(\cdot)$ above (we skip detailed steps for the sake of readability), we match the utility condition in \eqref{leftinvertibilitygeneral1} with the proposed affine maps in \eqref{privacymechanism2} which yields the following:
\begin{equation}\label{zkf}
   \tilde{g}(\tilde{\zeta}_k, \tilde{y}_k,w_k) := \Pi_3 {g}(\Pi_2^L \tilde{\zeta}_k,\Pi_1^L \tilde{y}_k,w_k)+ \Pi_4 \tilde{y}_{k},
\end{equation}
for full rank $\Pi_3 \in \mathbb{R}^{\tilde{n}_u \times n_u}$ and $\Pi_4 \in \mathbb{R}^{\tilde{n}_u \times \tilde{n}_y}$ to be designed. From this expression of $\tilde{g}(\cdot)$, it follows that $b_k^3 := \Pi_4\tilde{y}_k$ and $\tilde{u}_k$ is an affine function of $u_k$ given by
\begin{equation}\label{ukdistorted}
    \tilde{u}_k = \Pi_3 u_k + \Pi_4 \tilde{y}_{k}.
\end{equation}
The encoded data $\tilde{y}_{k}$ is used in $\tilde{g}(\cdot)$ to inject randomness into $\tilde{u}_k$. Finally, we design the extracting function $\pi_3^L(\cdot)$ satisfying the left invertibility condition, $\pi_3^L \circ \pi_3 (s) = s$. Given \eqref{ukdistorted}, the latter condition is written as
\begin{align} \label{left_inverse2}
\pi_3^L \left(\Pi_3 u_k + \Pi_4 \tilde{y}_{k}\right) =  u_k,
\end{align}
which trivially leads to
\begin{equation}
    \pi_3^L\left(\tilde{u}_k,\tilde{y}_k\right) := \Pi_3^L \left( \tilde{u}_k-\Pi_4 \tilde{y}_k\right). \label{inversemap}
\end{equation}
Matrix $\Pi_3^L$ always exists due to the rank of $\Pi_3$.

We can now state the proposed solution to Problem \ref{problem1}.

\begin{proposition}\label{proposition1}\emph{\textbf{(Solution to Problem \ref{problem1})}}
For given full rank matrices $\Pi_1 \in \mathbb{R}^{\tilde{n}_y \times n_y}$, $\Pi_2 \in \mathbb{R}^{\tilde{n}_\zeta \times n_\zeta}$, $\Pi_3 \in \mathbb{R}^{\tilde{n}_u \times n_u}$, and $\Pi_4 \in \mathbb{R}^{\tilde{n}_u \times \tilde{n}_y}$, matrix $N_1 \in \mathbb{R}^{\tilde{n}_y \times (\tilde{n}_y - n_y)}$ expanding the kernel of $\Pi_{1}^{L}$ \emph{(}i.e., $\Pi_{1}^{L}N_1=\mathbf{0}$\emph{)}, and random process $s_k \in \mathbb{R}^{(\tilde{n}_y - n_y)}$, the encoding map: \vspace{-1mm}
\begin{equation}\label{privacymechanisms}
\tilde{y}_k=\Pi_1 y_k + N_1 s_k,
\vspace{-1mm}
\end{equation}
target algorithm:
\begin{equation}\label{targetalgorithmsolution}
 \tilde{\Sigma}=\left\{\begin{array}{l}\begin{aligned}
\tilde{\zeta}_{k+1}&= \Pi_{2} f(\Pi_2^L \tilde{\zeta}_k,\Pi_1^L \tilde{y}_k,w_k), \\
\tilde{u}_k &= \Pi_3 {g}(\Pi_2^L \tilde{\zeta}_k,\Pi_1^L \tilde{y}_k,w_k)+ \Pi_4 \tilde{y}_{k},\end{aligned}
\end{array}\right.
\end{equation}
and inverse function:
\begin{equation}\label{inversesolution}
\pi_3^L\left(\tilde{u}_k,\tilde{y}_k\right) = \Pi_3^L \left( \tilde{u}_k-\Pi_4 \tilde{y}_k\right),
\end{equation}
provide a solution to Problem \ref{problem1}.
\end{proposition}
\emph{\textbf{Proof}}: Proposition \ref{proposition1} follows from the discussion provided in the solution section above, Section \ref{sec3}.
\hfill $\blacksquare$\\[1mm]
\begin{remark}
In Proposition \ref{proposition1}, we present the first part of the design of the proposed immersion-based coding mechanism for privacy. Note that the proposed scheme is independent of the nature of the user data $y_k$ and the utility $u_k$ and does not require any assumption on the original algorithm. These features make the proposed scheme suitable for enforcing privacy for a large class of linear and nonlinear high-dimensional algorithms.
\end{remark}
\begin{remark}
This manuscript primarily focuses on immersion-based coding for privacy of discrete-time algorithms operating in the cloud. It is noteworthy that it can be proved that the application of the proposed coding and target algorithm in Proposition \ref{proposition1} can be extended to privacy of continuous-time systems. This extension highlights the wider application of immersion-based coding, providing a solution for cloud-based privacy covering discrete and continuous-time algorithms.
\end{remark}
\begin{remark}
{In this section, we address Problem \ref{problem1} by imposing a specific structure on the encoding and immersion maps we aim to design. We focus on using random affine maps, which consist of linear coordinate transformations combined with additive random processes. This affine structure allows us to: 1) Develop a prescriptive solution to Problem \ref{problem1} based on the original algorithm functions $f(\cdot)$ and $g(\cdot)$. In other words, we leverage the original algorithm’s knowledge to construct the target algorithm algebraically; 2) Ensure that the immersion condition outlined in Problem \ref{problem1} is met; and 3) Utilize the additive random term—traceable through the algorithm and removable by the user—to enforce a desired level of differential privacy. It is important to note that this affine-based design approach for the encoding map and target algorithm is versatile, suitable for encoding various linear and nonlinear dynamical systems with different operators.}
\end{remark}
\subsection{Algorithms with Different Time-Scales}
The class of algorithms considered in \eqref{generaldynamics}, and the results that followed, operates on a single time scale. That is, the algorithm reacts to the received $y_k$, iterates once according to $f(\cdot)$ and $g(\cdot)$, and sends the utility $u_k$ back to the user. This is repeated sequentially between the user and the cloud. Algorithms that fit this class are, e.g., control, monitoring, and federated learning schemes. Note, however, that many algorithms operate on a different time scale from that of the user. That is, some algorithms receive the user data $y_k$ at time $k$, iterate locally multiple times in the cloud (say for $t=1,\ldots,T$), and only after some local iterations ($T$), the data utility $u_k$ is sent back to the user. Algorithms that fit this class are, for instance, general learning algorithms, where it is often the case that the complete data set is uploaded to the cloud, the training is done there by running gradient-like steps multiple times, and only when the cost function has decreased to an acceptable level, the final trained model is sent back to the user.\\
\indent For the sake of completeness, we concisely provide the corresponding coding result for algorithms with different time scales. Consider discrete-time dynamic algorithms of the form:
\begin{equation}\label{generaldynamics2Scales}
\Sigma =  \left\{\begin{array}{l}
\begin{aligned}
\zeta_{t+1}&=f\left(\zeta_t, y_k,w_t\right), \hspace{1mm} t=0,\ldots,T-1, \\
u_k &= g\left(\zeta_T,y_k,w_T \right),\\
\end{aligned}
\end{array}\right.
\end{equation}
with \emph{global} time-index $k \in \Nat_0$, \emph{local} time-index $t \in \Nat_0$, user data $y_k \in \mathbb{R}^{n_y}$, data utility $u_k \in \mathbb{R}^{n_u}$, algorithm internal variables $\zeta_t \in \mathbb{R}^{n_\zeta}$, unknown disturbance (or reference signal) $w_t \in \mathbb{R}^{n_w}$, and functions $(f,g)$ of appropriate dimensions.\\
\indent Further, consider the corresponding higher-dimensional target algorithm:
\begin{equation}\label{targetalgorithm2Scales}
\tilde{\Sigma} =  \left\{\begin{array}{l}\begin{aligned}
\tilde{\zeta}_{t+1}&=\tilde{f}(\tilde{\zeta}_t, \tilde{y}_k,w_t), \hspace{1mm} t=0,\ldots,T-1, \\
\tilde{u}_k&=\tilde{g}(\tilde{\zeta}_T, \tilde{y}_k,w_T),
\end{aligned}
\end{array}\right.
\end{equation}
with distorted user data $\tilde{y}_k \in \mathbb{R}^{\tilde{n}_y}$, $\tilde{n}_y > n_y$, distorted data utility $\tilde{u}_k \in \mathbb{R}^{\tilde{n}_u}$, $\tilde{n}_u > n_u$, algorithm variables $\tilde{\zeta}_t \in \mathbb{R}^{\tilde{n}_\zeta}$, $\tilde{n}_\zeta > n_\zeta$, and functions $(\tilde{f},\tilde{g})$ of appropriate dimensions.\\
\indent Following the reasoning described in the above section, for algorithm \eqref{generaldynamics2Scales} and corresponding target algorithm \eqref{targetalgorithm2Scales}, we let the user data encoding map, the immersion map, and the utility map be affine functions of the form:
\begin{equation}\label{maps2Scales}
    \left\{\begin{aligned}
    \tilde{y}_k &= \Pi_1 y_k + N_1 s_k, \\[-.5mm]
    \tilde{\zeta}_t &= \Pi_2 \zeta_t, \\[-.5mm]
    \tilde{u}_k  &= \Pi_3 u_k + \Pi_4\tilde{y}_k,
    \end{aligned}\right.
\end{equation}
for some full-rank matrices $(\Pi_1,\Pi_2,\Pi_3,\Pi_4,N_1)$ of appropriate dimensions. We can now state the problem we seek to address for \eqref{generaldynamics2Scales}-\eqref{maps2Scales}.

\begin{problem}\label{problem2}\emph{\textbf{(Privacy-Preserving Coding, Two Time Scales)}} For given $(f,g,\zeta_0)$ of the original algorithm \eqref{generaldynamics2Scales}, design an encoding matrix $\Pi_1$, random process $s_k$, and immersion matrix $\Pi_2$ in \eqref{maps2Scales}, and $(\tilde{f},\tilde{g},\tilde{\zeta}_0)$ of the target algorithm \eqref{targetalgorithm2Scales} so that: \textbf{(a)} $\tilde{\zeta}_t = \Pi_2 \zeta_t$ is forward invariant (the immersion condition); and \textbf{(b)} there exists a left-invertible matrix $\Pi_3$ and full rank matrix $\Pi_4$ satisfying $\tilde{u}_k = \Pi_3 u_k + \Pi_4\tilde{y}_k$.
\end{problem}

Problem \ref{problem2} is an analogue to Problem \ref{problem1} for systems with two time scales. Following the lines of the solution to Problem \ref{problem1}, we provide a solution to Problem \ref{problem2} in the following corollary of Proposition \ref{proposition1}.

\begin{corollary}\label{corollary1}\emph{\textbf{(Solution to Problem \ref{problem2})}}
For given full rank matrices $(\Pi_1,\Pi_2,\Pi_3,\Pi_4)$ of appropriate dimensions, matrix $N_1$ expanding the kernel of $\Pi_{1}^{L}$, and random process $s_k$, the encoding map: \vspace{-1mm}
\begin{equation}\label{privacymechanisms2scales}
\tilde{y}_k=\Pi_1 y_k + N_1 s_k,
\vspace{-1mm}
\end{equation}
target algorithm:
\begin{equation}\label{targetalgorithmsolution2scales}
\left\{\begin{array}{l}\begin{aligned}
\tilde{\zeta}_{t+1}&= \Pi_{2} f(\Pi_2^L \tilde{\zeta}_t,\Pi_1^L \tilde{y}_k,w_t), \hspace{1mm} t=0,\ldots,T-1,  \\
\tilde{u}_k &= \Pi_3 {g}(\Pi_2^L \tilde{\zeta}_T,\Pi_1^L \tilde{y}_k,w_T) + \Pi_4 \tilde{y}_{k},\end{aligned}
\end{array}\right.
\end{equation}
and inverse input map:
\begin{equation}\label{inversesolution2scales}
{u}_k = \Pi_3^L \left( \tilde{u}_k-\Pi_4 \tilde{y}_k\right),
\end{equation}
provide a solution to Problem \ref{problem2}.
\end{corollary}
\emph{\textbf{Proof}}: The proof of Corollary \ref{corollary1} follows the same lines as the proof of Proposition \ref{proposition1}. Only the time scales of the encoded signals change. The proof is omitted here for the sake of brevity.\\
\indent So far, we have presented the proposed immersion-based coding for two classes of discrete-time algorithms built around stochastic affine maps. Note that the only constraint on these maps is that they are full-rank. In the next section, we state sufficient conditions for the proposed privacy scheme to guarantee a prescribed level of differential privacy.
\section{Privacy Guarantees}\label{sec4}
\indent {The private element that we consider in the proposed scheme is privacy of the user data and utility. Even though we encode the original algorithm (which effectively makes inference of it from the encoded disclosed target algorithm difficult to accomplish), the algorithm itself is not considered private. The algorithm is encoded to make it work with encoded user data $\tilde{y}_k$ to produce encoded utility $\tilde{u}_k$.} Therefore, in what follows we focus on how to enforce differential privacy of $(y_k,u_k)$ by properly selecting the random process $s_k$ and the encoding matrices in the affine maps of user data $\tilde{y}_k = \Pi_1 y_k + N_1s_k$ and utility $\tilde{u}_k = \Pi_3 u_k + \Pi_4 \tilde{y}_{k}$.\\
\indent A privacy-level analysis of random affine maps of the form \eqref{privacymechanism2} has been discussed in the literature, see \cite{schluter2023cryptanalysis} (cryptanalysis) and \cite{upadhyay2013random} (DP). However, the guarantees they show are for square and invertible encoding matrices. In our scheme, we cannot directly apply those results as our encoding matrices ($\Pi_1$ and $\Pi_3$) are rectangular (tall matrices). We need rectangular matrices to be able to exploit their kernels to remove the injected distortion on the user side. A complication with this architecture is that the covariance matrix of $N_1s_k$ (which is part of the user data encoding mechanism) is degenerate regardless of the probability distribution of $s_k$. The latter prevents us from using off-the-shelf tools to design the affine maps to guarantee differential privacy.\\
\indent In what follows, we provide a tailored solution to guarantee DP for the class of mechanisms that we consider. In particular, we prove that the proposed scheme, with full-column rank encoding matrices, can provide any desired level of differential privacy without reducing the accuracy and performance of the original algorithm.
\subsection{{Threat Model and Security Analysis}}
{The immersion-based coding scheme is designed to maximize the difficulty of adversaries in trying to deduce private elements, including user data and utility. Our threat model assumes adversaries who can eavesdrop on data both within the communication network and in the cloud. These adversaries have access to all encoded input data $\tilde{y}_k$ and encoded utility $\tilde{u}_k$, and they are also aware of the specific target algorithm being executed in the cloud.\\
\indent In this scenario, a user aims to execute an algorithm remotely in the cloud. To apply the immersion-based coding scheme, the user first encodes the dynamic algorithm to create a target algorithm and then shares this encoded version with the cloud for remote execution. During each iteration, the user encodes input data and transmits them to the cloud, which then runs the target algorithm and produces the encoded utility. This utility is sent back to the user at each iteration and can be decoded to extract the original utility. Therefore in this setting, only the user can access the encoding matrices $\Pi_1$, $\Pi_2$, $\Pi_3$, and $\Pi_4$. The cloud only has access to the target algorithm $\tilde{\Sigma}$, and does not have access to the original algorithm $\Sigma$. Therefore, adversaries at the cloud cannot derive the encoding matrices $\Pi_1$, $\Pi_2$, $\Pi_3$, and $\Pi_4$ from the functions $\tilde{f}(\cdot)$ and $\tilde{g}(\cdot)$ in the target algorithm, as these functions cannot be compared with $f(\cdot)$ and $g(\cdot)$ from the original algorithm. Consequently, the cloud does not need to be trusted, since it only interacts with encoded data that only the user can decode.\\
\indent In the following sections, we will demonstrate how this immersion-based coding scheme can provide the desired differential privacy guarantees for the private elements of remote dynamic algorithms.}
\subsection{Differential Privacy}
In the context of databases, $(\epsilon, \delta)$-Differential Privacy (DP) \cite{Dwork} was introduced as a probabilistic framework to quantify privacy of probabilistic maps. The constant $\epsilon \geq 0$ quantifies how similar (different) the output of a mechanism is in \emph{adjacent} datasets, say $(\mathcal{D}$ and $\mathcal{D}^{\prime})$, and $\delta$ is a constant shift used when the ratio of the probabilities of $\mathcal{D}$ and $\mathcal{D}^{\prime}$ under the mechanism cannot be bounded by $e^\epsilon$ (see Definition \ref{definition2} below). With an arbitrarily given $\delta$, a mechanism with a smaller $\epsilon$ makes \emph{adjacent} data less distinguishable and hence more private.

\begin{definition}\label{definition1}[Adjacency]
Let $\mathcal{X}$ denote the space of all possible datasets. We say that $\mathcal{D} \in \mathcal{X}$ and $\mathcal{D}^{\prime} \in \mathcal{X}$ are adjacent if they differ on a single element.
\end{definition}

\begin{definition}\label{definition2} [$(\epsilon,\delta)$-Differential Privacy \cite{Dwork2}]\label{DP1} The random mechanism $\mathcal{M}: \mathcal{X} \rightarrow \mathcal{R}$ with domain $\mathcal{X}$ and range $\mathcal{R}$ is said to provide $(\epsilon,\delta)$-differential privacy, if for any two adjacent datasets $\mathcal{D}, \mathcal{D}^{\prime} \in \mathcal{X}$ and for all measurable sets $\mathcal{S} \subseteq \mathcal{R}$\emph{:}
\begin{equation}\label{DP}
\operatorname{Pr}(\mathcal{M}(\mathcal{D}) \in  \mathcal{S}) \leq e^\epsilon \operatorname{Pr}\left(\mathcal{M}\left(\mathcal{D}^{\prime}\right) \in  \mathcal{S}\right)+\delta.
\end{equation}
\end{definition}
If $\delta=0$, $\mathcal{M}$ is said to satisfy $\epsilon$-differential privacy. From Definition \ref{definition2}, we have that a mechanism provides DP if its probability distribution satisfies \eqref{DP} for some $\epsilon$ and $\delta$. Then, if we seek to design the mechanism to guarantee DP, we need to shape its probability distribution. This is usually done by injecting noise into the data we seek to encode. The noise statistics must be designed in terms of the sensitivity of the data to be encoded. Sensitivity refers to the maximum change of the data due to the difference in a single element of the dataset.
\begin{definition} [Sensitivity]\label{Sensitivity}\emph{:} Given two adjacent datasets $\mathcal{D}, \mathcal{D}^{\prime} \in \mathcal{X}$, and a query function $q: \mathcal{X} \rightarrow \mathcal{R}$ (a deterministic function of datasets). The sensitivity of $q(\cdot)$ is formulated as
\begin{equation}\label{sensitivity}
  \Delta^q_1=\sup _{\mathcal{D}, \mathcal{D}^{\prime}}\left\|q\right(\mathcal{D} \left) - q\left(\mathcal{D}^{\prime} \right)\right\|_1.
\end{equation}
\end{definition}
A differential privacy mechanism $\mathcal{M}$ based on the sensitivity \eqref{sensitivity} needs to be designed so that the differential privacy condition \eqref{DP} holds.

\subsection{Immersion-based Coding Differential Privacy Guarantee}
We formulate the problem of designing the variables of the privacy mechanisms in \eqref{privacymechanisms} and \eqref{targetalgorithmsolution} (or \eqref{privacymechanisms2scales} and \eqref{targetalgorithmsolution2scales} for the two time scales algorithm) that, at every time step, guarantee differential privacy of the user and utility data $(y_k,u_k)$ from the disclosed distorted data in the cloud, $(\tilde{y}_k,\tilde{u}_k)$. We seek to design matrices $\Pi_1$, $\Pi_3$, and $\Pi_4$, and a random process $s_k$ such that $\tilde{y}_k$ and $\tilde{u}_k$ are $\epsilon^y$ and $\epsilon^u$-differentially private, respectively.
\begin{problem}\label{problem3}\emph{\textbf{(Element-Wise Differential Privacy)}}  Given the desired privacy levels $\epsilon^y$ and $\epsilon^u$, design the variables of privacy mechanisms in \eqref{privacymechanisms} and \eqref{targetalgorithmsolution} (or \eqref{privacymechanisms2scales} and \eqref{targetalgorithmsolution2scales} for the two time scales algorithm) such that at time $k$, each element of $\tilde{y}_k$ and $\tilde{u}_k$, $\tilde{y}^i_k$ and $\tilde{u}^j_k$, $i \in \{1,...,\tilde{n}_y \}$ and $j \in \{1,...,\tilde{n}_u\}$, are $\epsilon^y$ and $\epsilon^u$-differentially private, respectively, for any measurable $\mathcal{S}_y,\mathcal{S}_u \subset \mathbb{R}$, i.e.,
\begin{equation}\label{dpcondition}
    \left\{\begin{aligned}
\mathbb{P}( \tilde{y}^i_{k} (y_k)\in \mathcal{S}_y) &\leq e^{\epsilon^y} \mathbb{P}(\tilde{y}^i_{k} (y^{\prime}_k)\in \mathcal{S}_y),\\[1mm]
\mathbb{P}( \tilde{u}^j_{k} (u_k) \in \mathcal{S}_u) &\leq e^{\epsilon^u} \mathbb{P}(\tilde{u}^j_{k} (u^{\prime}_k)\in \mathcal{S}_u), \\[1mm]
&\hspace{-23mm} \text {for adjacent}\left(y_k,y^{\prime}_k\right) \text { and }\left(u_k,u^{\prime}_k\right).
    \end{aligned}\right.
\end{equation}
\end{problem}
\subsection*{Solution to Problem \ref{problem3}}
As standard in the DP literature, we let the stochastic process $s_k$ follow a multivariate i.i.d. Laplace distribution with mean $E[s_k] =: \mu \in \Real^{\tilde{n}_y - n_y}$ and covariance matrix $E[(s_k-\mu)(s_k-\mu)^\top] =: \sigma I_{(\tilde{n}_y - n_y)}$, for some $\sigma>0$, i.e., $s_k \sim \operatorname{Laplace }(\mu, \sigma I_{(\tilde{n}_y - n_y)})$.\\
\indent We start with the privacy guarantee for $\tilde{y}_k$. According to Definition \ref{Sensitivity}, given adjacent user data $y_k,y^{\prime}_k \in \mathcal{Y}$, where $\mathcal{Y}$ denotes the space of all user data sets, the sensitivity of $y_k$ is as follows:
\begin{equation}\label{sensitivityy}
    \Delta_1^y=\sup _{y_k,y^{\prime}_k \in \mathcal{Y}} \left\|y_k - y^{\prime}_k\right\|_1.
\end{equation}
Because $s_k \sim \operatorname{Laplace }(\mu, \sigma I)$, and given the privacy encoding mechanisms \eqref{privacymechanisms} or \eqref{privacymechanisms2scales}, each element of $\tilde{y}_k$ also follows a Laplace distribution:
\begin{align}
\tilde{y}_{k}^{i} \sim \operatorname{Laplace}\left(N_1^{i} \mu+\Pi_{1}^{i} y_{k},||N_1^{i}||_2 \sigma \right),
\end{align}
where $N_1^{i}$ and $\Pi_1^{i}$ are the $i$-th rows of matrices $N_1$ and $\Pi_1$, respectively. It follows that 
\begin{equation}\label{ineq1dp}
\begin{aligned}
    &\mathbb{P}\left(\tilde{y}^i_{k}(y_k) \in \mathcal{S}_y\right) =\left(\tfrac{1}{2  \left\|N_1^{i}\right\|_2 \sigma}\right) \int_{\mathcal{S}_y}  e^{\frac{-\left\|p-\Pi_{1}^{i} y_{k}-N_1^{i} \mu \right\|_{1}}{ \left\|N_1^{i}\right\|_2 \sigma}} d p\\
    &\stackrel{(\mathrm{a})}{\leq}  e^{\frac{\left\|\Pi_{1}^{i}\left({y}_k -y^{\prime}_k \right)\right\|_1}{\left\|N_1^{i}\right\|_2 \sigma}}  \left( \tfrac{1}{2 \left\|N_1^i\right\|_2 \sigma} \right) \int_{\mathcal{S}_y} e^{\frac{-\left\|p-\Pi_{1}^{i} y^{\prime}_k -N_1^{i} \mu\right\|_{1}}{ \left\|N_1^{i}\right\|_2 \sigma}}d p\\&= e^{\frac{\left\|\Pi_{1}^{i}\left({y}_k -y^{\prime}_k\right)\right\|_1}{\left\|N_1^{i}\right\|_2 \sigma }} \mathbb{P}\left( \tilde{y}^i_{k}(y_k^\prime) \in \mathcal{S}_y\right),
    \end{aligned}
\end{equation}
where (a) follows from the following triangle inequality
\begin{equation}
\begin{aligned}
    &-\left\|p -\Pi_{1}^{i} {y}_k -N_1^{i} \mu\right\|_{1} \leq -\left\|p-\Pi_{1}^{i} y^{i\prime}_k-N_1^{i} \mu\right\|_{1}\\[1mm]&\hspace{45mm}+\left\|\Pi_{1}^{i}\left({y}_k -y^{i\prime}_k\right)\right\|_1.
    \end{aligned}
    \end{equation}
Due to the sensitivity relation \eqref{sensitivityy}, we have
\begin{equation}
\left\|\Pi_{1}^{i}\left({y}_k - y^{\prime}_k \right)\right\|_1 \leq \left\|\Pi_{1}^{i}\right\|_1 \Delta_1^y.
\end{equation}
Hence, inequality \eqref{ineq1dp} implies 
\begin{equation}
        \mathbb{P}\left(\tilde{y}^i_{k}(y_k)  \in \mathcal{S}_y\right) \le e^{\frac{\left\|\Pi_{1}^{i}\right\|_1 \Delta_1^y}{ \left\|N_1^{i}\right\|_2 \sigma}} \mathbb{P}\left(\tilde{y}^i_{k}(y_k^\prime) \in \mathcal{S}_y\right).
\end{equation}
Therefore, $\epsilon^y$-differential privacy of $y^i_{k}$ given $\tilde{y}_{k}$, for all $i \in \{1,...,\tilde{n}_y\}$, is guaranteed for $\Pi_1$, $N_1$, and $\sigma$ satisfying 
\begin{equation}\label{dpnoisey}
    \frac{||\Pi_{1}^{i}||_1 \Delta_1^y}{ ||N_1^{i}||_2 \sigma} \le {\epsilon^y}.
\end{equation}
Following the same reasoning, it can be shown that $u^j_{k}$ is $\epsilon^u$-differentially private for given $\tilde{u}_{k}$ and all $j \in \{1,...,\tilde{n}_u\}$, if $\Pi_3$, $\Pi_4$, $N_1$, and $\sigma$ satisfy
\begin{equation}\label{dpnoiseua}
    \frac{||\Pi_{3}^{j}||_1 \Delta_1^u}{ ||\Pi_4 N_1^{j}||_2 \sigma}  \le {\epsilon^u},
\end{equation}
where $\Pi_3^{j}$ is the $j$-th row of matrix $\Pi_3$,
\begin{equation}\label{sensitivityu}
    \Delta_1^u = \sup _{u_k,u^{\prime}_k \in \mathcal{U}} \left\|u_k - u^{\prime}_k\right\|_1,
\end{equation}
and $\mathcal{U}$ denoting the utility space.\\
\indent Hence, according to the differential privacy conditions, \eqref{dpnoisey} and \eqref{dpnoiseua}, to increase the privacy guarantees (by decreasing $\epsilon^y$ and $\epsilon^u$), we need to design $\Pi_1$ and $\Pi_3$ as small as possible while selecting $\sigma$ (the covariance of $s_k$) and $\Pi_4$ as large as possible. From \eqref{privacymechanisms}, \eqref{targetalgorithmsolution}, \eqref{privacymechanisms2scales}, and \eqref{targetalgorithmsolution2scales}, it is apparent that by choosing small $\Pi_1$ and $\Pi_3$, and large $\sigma$ and $\Pi_4$, $\tilde{y}_k$ and $\tilde{u}_k$ are close to $N_1s_k$ and $\Pi_4 N_1s_k$, respectively, and (practically) independent from ${y}_k$ and ${u}_k$. Note that $\left\|N_1^{i}\right\|_2$, $i \in \{1,\dots,\tilde{n}_y \}$, must be nonzero, i.e., we need to design $N_1$ without zero rows. The latter is not a technical constraint as, for a given $N_1$ with nonzero rows, $\Pi_1$ can be obtained by solving the equation $\Pi_1^L N_1=\mathbf{0}$ and computing the right inverse of $\Pi_1^L$.\\
\indent {The conditions on $\Pi_1$, $\Pi_3$, $\Pi_4$, $N_1$, and $\sigma$ to guarantee the desired level of privacy are provided in the following theorem.
\begin{theorem}\label{theoremLaplace}\emph{\textbf{(Differential Privacy through Laplace additive noises)}}
Consider the mechanisms \eqref{ukdistorted} and \eqref{privacymechanisms} (or \eqref{maps2Scales} for the two time scales algorithm) with full-rank matrices $\Pi_1 \in \mathbb{R}^{\tilde{n}_y \times n}$, $\Pi_3 \in \mathbb{R}^{\tilde{n}_u \times n_u}$, $\Pi_4 \in \mathbb{R}^{\tilde{n}_u \times n_y}$, matrix $N_1 \in \mathbb{R}^{\tilde{n}_y \times (\tilde{n}_y - n_y)}$ expanding the kernel of $\Pi_{1}^{L}$, and stochastic Laplace process $s_k \sim \operatorname{Laplace }(\mu, \sigma I)$ with standard deviation $\sigma$. If the following inequalities are satisfied: 
\begin{equation}\label{ineq laplace all}
\left\{ \begin{aligned}
&\frac{||\Pi_{1}^{i}||_1 \Delta_1^y}{ ||N_1^{i}||_2 \sigma} \le {\epsilon^y},\\
&\frac{||\Pi_{3}^{j}||_1 \Delta_1^u}{ ||\Pi_4 N_1^{j}||_2 \sigma}  \le {\epsilon^u},
\end{aligned}\right.
\end{equation}
where $\Delta_1^y$ and $\Delta_1^u$ are the sensitivities of $y_k$ and $u_k$, respectively, as defined in \eqref{sensitivityy} and \eqref{sensitivityu}, then, the mechanisms \eqref{ukdistorted} and \eqref{privacymechanisms} (or \eqref{maps2Scales} for the two time scales algorithm) provide ${\epsilon^y}$ and ${\epsilon^u}$- Differential Privacy Guarantee for each element of vectors $\Tilde{y}_k$ and $\Tilde{u}_k$, $\Tilde{y}^i_k$ and $\Tilde{u}^j_k$, respectively, for $i \in \{1,...,\Tilde{n}_y \}$ and $j \in \{1,...,\tilde{n}_u\}$ (in the sense of Problem \ref{problem3}).
\end{theorem}}
%
\subsection{Perfect Security}
A perfectly secure cryptosystem is a secure system against adversaries with unlimited computing resources and time; see \cite{diffie2019new}. In \cite{shannon1949communication}, Shannon proves that a necessary condition for an encryption method to be perfectly secure is that the uncertainty of the secret key is larger than or equal to the uncertainty of the plaintext; see \cite{wang2008book}. He proposes a one-time pad encryption scheme in which the key is randomly selected and never used again. The one-time pad gives unbounded entropy of the key space, i.e., infinite key space, which provides perfect security. Perfect secrecy is lost when the key is not random or is reused. This perfect security idea can be used to assess the performance of our proposed coding mechanism. Given the fact that in the proposed mechanism, \eqref{privacymechanism2}, the encoding keys, $(\Pi_1,s_k)$ and $(\Pi_3,\Pi_4 N_1 s_k)$, are random, change in every iteration, and provide an infinite-dimensional key space (the support of $s_k$), can be considered perfectly secure.\\
\indent Perfect secrecy of the proposed coding mechanisms can also be concluded from the differential privacy point of view. In \cite{dwork2011differential}, the authors define perfect secrecy in terms of DP. They state that a mechanism $\mathcal{M}: \mathcal{X} \rightarrow \mathcal{R}$, with domain $\mathcal{X}$ and range $\mathcal{R}$ is perfectly secret if
\begin{equation}\label{perfectsecrecy}
\frac{\operatorname{Pr}(M(\mathcal{D}) \in \mathcal{S})}{  \operatorname{Pr}\left(M\left(\mathcal{D}^{\prime}\right) \in \mathcal{S}\right)}=1=e^0,
\end{equation}
for any two adjacent $\mathcal{D},\mathcal{D}^{\prime} \in \mathcal{X}$ and for all measurable sets $\mathcal{S} \subseteq \mathcal{R}$. Then, the output of the mechanism reveals no information about private data. Comparing the definition of perfect secrecy in \eqref{perfectsecrecy} and the definition of differential privacy (Definition \ref{definition2}), we can conclude that a differentially private mechanism is perfectly secret if $(\epsilon,\delta)$ equal zero.\\
\indent The relation between design variables of the proposed encoding mechanism and its differential privacy guarantees is shown in \eqref{dpnoisey} and \eqref{dpnoiseua}. Therefore, to have perfect secrecy, we need $(\epsilon^y,\epsilon^u) = \mathbf{0}$. While this can not be achieved, $(\epsilon^y, \epsilon^u)$ can be made arbitrarily small by selecting small $\Pi_1$ and $\Pi_2$ and large $\sigma$ and $\Pi_4$.

\section{General Guidelines for Implementation}\label{sec5}
In what follows, we provide a synthesis procedure to summarize the design steps and operation methodology for the proposed mechanism. 
\noindent\makebox[\linewidth]{\rule{\linewidth}{0.8pt}}
\textbf{Synthesis and Operation Procedure:}\\
{\begin{itemize}
    \item The user randomly selects the "small" full rank matrices $(\Pi_1,\Pi_2,\Pi_3)$, and the "large" full rank matrix $\Pi_4$ of appropriate dimensions (see Proposition \ref{proposition1}) and computes the left inverse matrices $\Pi^L_1$ and $\Pi^L_2$, and base $N_1$ of the kernel of $\Pi_{1}^{L}$. User fixes the distribution of the random process $s_k$ in \eqref{privacymechanisms}. These matrices are designed to satisfy the desired level of DP guarantee based on the condition given in Theorem \ref{theoremLaplace}.\vspace{2mm}
    \item User creates the target algorithm \eqref{targetalgorithmsolution} (or \eqref{targetalgorithmsolution2scales} for the two time scales algorithm) based on its original dynamical algorithm and designed mechanism matrices and sends the target algorithm to the cloud.\vspace{2mm}
    \item At every time step $k \in \Nat_0$, user encodes $y_k$ according to \eqref{privacymechanisms} and sends $\tilde{y}_k$ to the cloud.\vspace{2mm}
    \item The cloud runs the target algorithm in \eqref{targetalgorithmsolution} (or \eqref{targetalgorithmsolution2scales} for the two time scales algorithm) using the received encoded $\tilde{y}_k$ to create the encoded utility $\tilde{u}_k$.\vspace{2mm}
    \item The cloud sends the encoded utility $\tilde{u}_k$ back to the user.\vspace{2mm}
    \item The user decodes the encoded utility signal using the inverse function in \eqref{inversesolution} to extract the original utility.
\end{itemize}}
\noindent\makebox[\linewidth]{\rule{\linewidth}{0.8pt}}
\section{Case Studies}\label{sec6}
In this section, we show the performance of the immersion-based coding algorithm on two use cases -- privacy in optimization/learning algorithms and privacy in a nonlinear networked control system.
\subsection{Immersion-based Coding for Privacy in Optimization/Learning algorithms}
\indent Machine learning (ML) has been successfully used in a wide variety of applications for multiple fields and industries \cite{jordan2015machine}, including healthcare, vehicular networks and intelligent manufacturing \cite{shinde2018review}.
However, collecting large amounts of training data poses a challenge for ML systems. Typically, increased data leads to improved ML model performance, necessitating the need for large datasets, often sourced from various origins. However, the process of gathering and using such data, along with the development of ML models, presents significant privacy risks due to the potential exposure of private information.\\
\subsubsection{Related work}
{In recent years, various privacy-preserving schemes have been implemented to address the privacy leakage in ML. Most of them rely on perturbation-based techniques such as Differential Privacy (DP) \cite{wei2020federated}, and cryptography-based techniques such as Secure Multiparty Computation (SMC) \cite{shokri2015privacy} and Homomorphic Encryption (HE) \cite{aono2017privacy}.\\
\indent Differential Privacy (DP) \cite{abadi2016deep} is a popular tool to enforce privacy for machine learning. The idea of training ML models with DP guarantee is proposed in \cite{abadi2016deep} (DP-SGD). This method is an adaptation of Stochastic Gradient Descent (SGD) that clips individual gradients and adds Gaussian noise to their average. Although DP provides strong information-theoretic guarantees and has a small system overhead, existing DP techniques for ML impose a large degradation to the utility of the trained models compared to non-private models. Recent works have used various techniques to improve the utility-privacy trade-off of such private ML techniques (e.g., by scaling the hyper-parameters \cite{sander2023tan}); however, there still exists a huge gap between the accuracy of DP-guaranteeing ML mechanisms and their non-private alternatives.\\
\indent Homomorphic Encryption (HE) is another prevalent technique for Privacy-Preserving Machine Learning (PPML), which allows performing a series of linear operations directly on encrypted data. The initial work on Privacy-Preserving Neural Networks (PPNN) with provable security was conducted by Barni et al. \cite{barni2006privacy} using additive HE and garbled circuits. Following the advent of Fully Homomorphic Encryption (FHE), several PPML techniques, including CryptoNet \cite{gilad2016cryptonets} and CryptoDL \cite{hesamifard2017cryptodl}, have been proposed. These techniques simulate neural network computations on encrypted data by deriving approximate polynomials for nonlinear activation functions. However, achieving accuracy in nonlinear layers requires high-order polynomial approximations, resulting in significant overhead. Furthermore, not all nonlinear functions can be effectively approximated using polynomials. Some methods leverage Secure Multiparty Computation (SMC) to compute nonlinear layers, as discussed by Shokri et al. \cite{shokri2015privacy}, but SMC is less attractive to cloud providers due to the need for model sharing with clients and the substantial computational and communication overhead involved. Word-wise FHE schemes, such as the Cheon-Kim-Kim-Song (CKKS) scheme \cite{cheon2017homomorphic}, are more suitable for PPML because they can handle encrypted real numbers. Nevertheless, they still only support homomorphic addition and multiplication, and not common non-arithmetic activation functions like ReLU and sigmoid. Replacing these functions with simple polynomials has not yielded high accuracy for advanced tasks and leads to performance degradation in ML models. Despite considerable advancements, FHE-based PPML remains computationally intensive \cite{liu2021machine}.\\ 
\indent In conclusion, while current solutions enhance the privacy of ML algorithms, they often compromise model performance and system efficiency. In this paper, we propose a PPML framework based on the immersion-based coding scheme outlined in Corollary \ref{corollary1}. We evaluate the performance of the proposed System Immersion-based privacy-preserving ML (SIML) scheme by comparing it with two well-known privacy-preserving ML tools, DP-SGD and CKKS, in terms of computational complexity and network accuracy. The results demonstrate that our framework outperforms these privacy-preserving tools by maintaining the same accuracy and convergence rate as standard machine learning, being computationally efficient, and offering any desired level of differential privacy.} 
\subsubsection{Learning Model Parameters using Gradient Descent}
Learning the parameters of a machine learning model is a nonlinear optimization problem. The algorithms used to solve these problems are typically variants of gradient descent \cite{ruder2016overview}. 
Simply put, gradient descent starts at a random point (set of parameters for the model), then, at each step, computes the gradient of the nonlinear function being optimized, called the loss function, and updates the parameters so as to decrease the gradient. This process continues until the algorithm converges to a local optimum. Hence, the vector of optimal model parameters is obtained by $w^*=\arg \min _{w} l\left({w},\mathcal{D}\right)$, with the database $\mathcal{D}$ of size $n_D$, the model parameter vector $w$, and the loss function $l\left({w},\mathcal{D}\right)$.\\
\indent In gradient descent optimization algorithms, the model parameters are updated following $T$ iterations of the optimizers that can generally be modeled as follows:
\begin{equation}
\text{Optimizer} \left\{
\begin{aligned}
&w_{{t+1}} = {w}_{t} -
p({w}_{t},\mathcal{D}),  \label{standardoptimizer}\\[1mm]
&t=0,1, \cdots, T-1,\\[1mm]
&w^{*} = w_{T},
\end{aligned}\right.
\end{equation}
where $w_{{t}} \in \mathbb{R}^{n_w}$ denotes the model parameters vector at the $t$-th epoch of the optimization algorithm, $T$ is the total number of epochs, and $p(\cdot,\cdot)$ is a gradient-based function that can be defined for each gradient descent optimizer, separately.\\
Stochastic Gradient Descent (SGD) is one of the most common gradient descent optimization techniques in machine learning \cite{ruder2016overview}. For SGD, function $p(\cdot,\cdot)$ can be written as follows:
\begin{equation}
p({w}_{t},\mathcal{D}):=\eta \nabla {l}({w}_{t},\mathcal{Y}_t),
\end{equation}
where $\eta>0$ is learning rate, $\nabla {l}(\cdot,\cdot)$ is the gradient of loss function ${l}(\cdot,\cdot)$, and $\mathcal{Y}_t$ is a mini-batch of database (a subset of database $\mathcal{D}$) for t-th epoch.
Another well-known gradient descent optimizer in ML is the Adam optimization algorithm \cite{ruder2016overview}. Adam is an extended version of stochastic gradient descent implemented in various ML applications. Function $p(\cdot,\cdot)$ can be modeled for Adam optimizer as follows:
\begin{equation}\label{adammain}
\text{(Adam)} \left\{
\begin{aligned}
&m_{t+1}=\beta_1 m_{t}+\left(1-\beta_1\right) \nabla l\left(w_{t}, \mathcal{Y}_t\right), \\
&v_{t+1}=\beta_2 v_{t}+\left(1-\beta_2\right)\left(\nabla l\left(w_{t}, \mathcal{Y}_t\right)\right)^2, \\
&p({w}_{t},\mathcal{D})=\frac{\frac{\alpha}{1-\beta_1^t} m_{t+1}}{\sqrt{\frac{v_{t+1}}{1-\beta_2^t}}+\epsilon},
\end{aligned}\right.
\end{equation}
where $m_{{t}} \in \mathbb{R}^{n_w}$, and $v_{{t}} \in \mathbb{R}^{n_w}$ denote the aggregate of gradients and the sum of the square of past gradients, respectively, in the $t$-th epoch of the Adam algorithm, $\alpha>0$ is the step size, $\beta_1, \beta_2 \in[0,1)$ are exponential decay rates for the moment estimates, $\epsilon>0$ is a small constant to avoid division by zero, and $\mathcal{Y}_t$ is a mini-batch of database (a subset of database $\mathcal{D}$) for t-th epoch.
\subsubsection{Immersion-based Coding for Privacy-Preserving Gradient Descent Optimizers}
In this section, we encode the gradient descent optimization algorithm given in \eqref{standardoptimizer}, following the same technique that we discussed in Corollary \ref{corollary1}, which is the immersion-based coding for remote dynamical algorithms with different time scales. The encoded optimization algorithm can be employed to provide privacy for the database and the model in machine learning algorithms.\\
\indent In this application, the optimizer in \eqref{standardoptimizer} is the original dynamical system $\Sigma$ in \eqref{generaldynamics2Scales} that we want to immerse to provide privacy. We consider a setting where the user who owns the database uploads the complete database $\mathcal{D}$ to the cloud, where the training process is executed by running the gradient-based optimizer \eqref{standardoptimizer}. The optimizer iterates $T$ times in the cloud to converge to a local optimum. Subsequently, the optimal model parameter vector $w^*$ is sent back to the user. Since the user and the cloud send the database and the optimal model, which play the role of input and utility of the algorithm, only once, the number of user iterations is $K=1$, and the number of local iterations is $T$. Functions $f(\cdot)$ and $g(\cdot)$ in \eqref{generaldynamics2Scales} can be modeled as $w_{t+1}=f(w_{t},\mathcal{D}):={w}_{t} -
p({w}_{t},\mathcal{D})$ and $w^*=g(w_{T}):=w_T$ in this case. The model parameters $w_{t}$, the database $\mathcal{D}$, and the vector of optimal model parameters $w^*$ can be considered as the internal variables $\zeta_t$, input $y$, and utility $u$ of the original algorithm $\Sigma$ in \eqref{generaldynamics2Scales}.\\
\indent To implement the user data encoding map in \eqref{maps2Scales} for the user database $\mathcal{D}$, we assume that the user distorts each data of the database $\mathcal{D}$, $y_i \in \mathbb{R}^{{n}_y}$, with $i=\{1,2,...,n_D\}$, before sending it to the cloud, as follows:
\begin{equation}\label{distortopt1}
\Tilde{y}_i=\pi_1\left(y_i\right)=\Pi_1 y_{i} + N_1 s_i, \forall y_i \in \mathcal{D},
\end{equation}
for full column rank matrix $\Pi_1 \in \mathbb{R}^{\Tilde{n}_y \times n_y}$, $N_1 \in \mathbb{R}^{\Tilde{n}_y \times (\Tilde{n}_y-{n}_y)}$ expanding the kernel of $\Pi_1^L$, random vectors $s_i \in \mathbb{R}^{(\Tilde{n}_y-{n}_y)}$, and distorted data $\Tilde{y}_i \in \mathbb{R}^{\tilde{n}_y}$ with $i=\{1,2,...,n_D\}$. Then, user sends the distorted database $\tilde{\mathcal{D}}$, which includes all distorted data $\Tilde{y}_i$, $i \in \{1,...,n_D \}$, to the cloud.\\
We use the distortion and immersion maps and the target dynamical algorithm in Corollary \ref{corollary1} to immerse the model parameters and optimization algorithm as follows.

The immersion and utility encoding maps are given by:
\begin{equation}\label{distortopt2}
\left\{\begin{array}{l}
 \tilde{w}_{t}= \Pi_2 {w}_{t},\\
 \tilde{w}^*=\Pi_3{w}^*+\Pi_4 \Tilde{y}_0,
 \end{array}\right.
\end{equation}
the target gradient descent optimizer is given by:
\begin{equation}\label{targetoptimizer}
 (\text {Target Optimizer})\left\{\begin{array}{l}
\tilde{w}_{{t+1}}=\Tilde{f}(\tilde{w}_{t},\tilde{\mathcal{D}})\\
\,\,\,\,\,\,\,\,\,\,\,\,\,:=\tilde{w}_{t}-\Pi_2 p\left(\Pi_2^L \tilde{w}_{t}, \bar{\mathcal{D}}\right), \\
t=0,1, \cdots, T-1, \\
\tilde{w}^*=\tilde{g} (\tilde{w}_{T},\tilde{\mathcal{D}}):=\Pi_3 {\Pi}_2^L \tilde{w}_{T}+\Pi_4 \Tilde{y}_0,
\end{array}\right.
\end{equation}
and the inverse function:
\begin{equation}
\pi_3^L(\tilde{w}^*,\tilde{\mathcal{D}}) = \Pi_3^L (\tilde{w}^* - \Pi_4 \Tilde{y}_0),
\end{equation}
with full rank matrices $\Pi_2 \in \mathbb{R}^{\Tilde{n}_w \times {n}_w}$, $\Pi_3 \in \mathbb{R}^{\Tilde{n}_w \times n_w}$, and $\Pi_4 \in \mathbb{R}^{\Tilde{n}_w \times \Tilde{n}_y}$ and the decoded database $\bar{\mathcal{D}}$ with $\bar{y}_i = \Pi_1^L  \tilde{y}_i \in \bar{\mathcal{D}}$, $i=\{1,2,...,n_D\}$.

This approach can be used to encode all kinds of gradient descent optimization algorithms that are typically used by machine learning algorithms. We will refer to machine learning algorithms that utilize immersion-based coding to preserve privacy as System Immersion-based Machine Learning (SIML).
\begin{figure}[!htb]
\centering
\subfloat[Original Image]{\includegraphics[width = 1.6in]{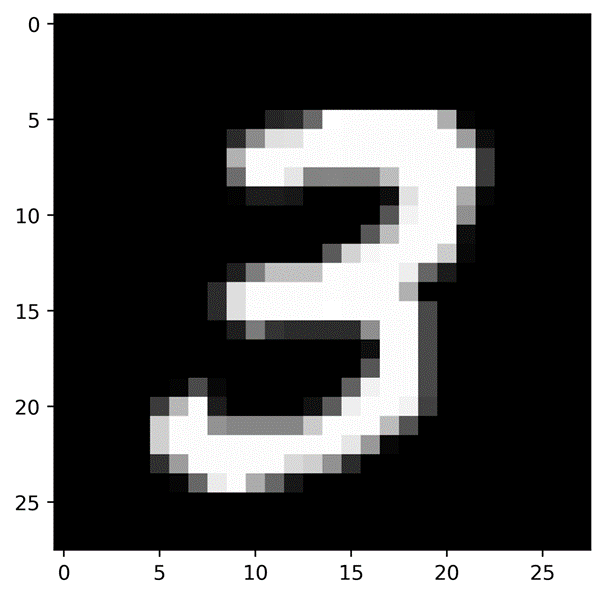}}
\subfloat[Encoded Image]{\includegraphics[width = 1.6in]{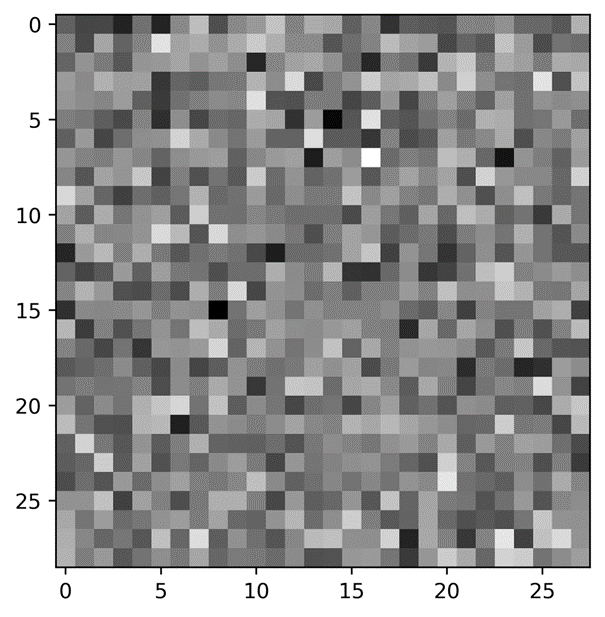}}
\caption{The comparison of one sample of the original MNIST database and its encoded format.}
\label{fig:encodedimage}
\end{figure}
\subsubsection{Case Study of Optimization/Learning algorithms}
In this section, we implement our proposed SIML scheme for performance evaluation using a multi-layer perception (MLP) \cite{tang2015extreme} and a real-world machine-learning database. Our investigation involves utilizing two optimization tools, i.e., Adam and SGD, on the MNIST database \cite{lecun1998gradient}. The experimental details are described as follows:
\begin{figure*}[t]
\centering
  \subcaptionbox{}[.4\linewidth][c]{%
    \includegraphics[width=1\linewidth]{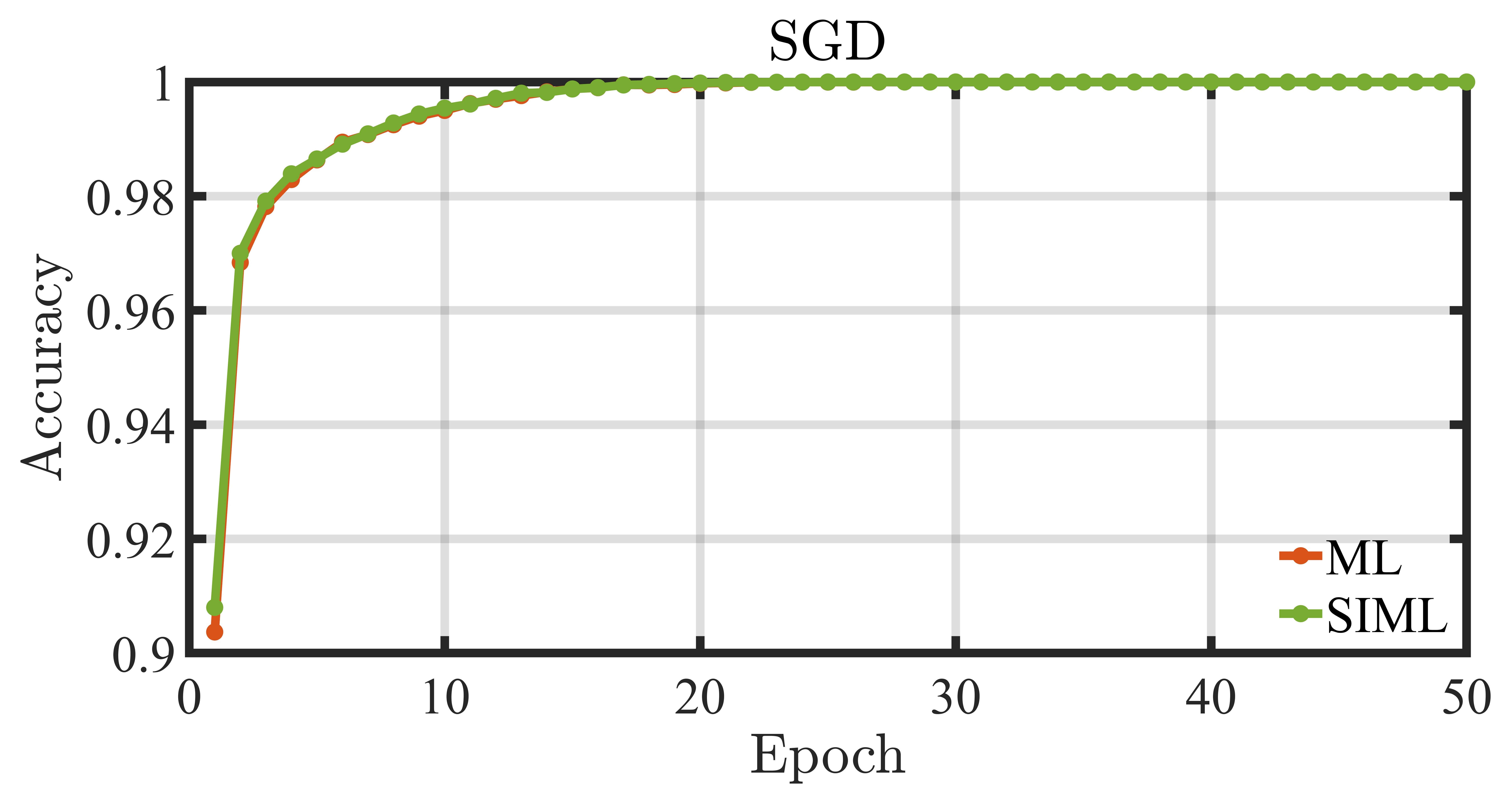}}
\quad
  \subcaptionbox{}[.4\linewidth][c]{%
    \includegraphics[width=1\linewidth]{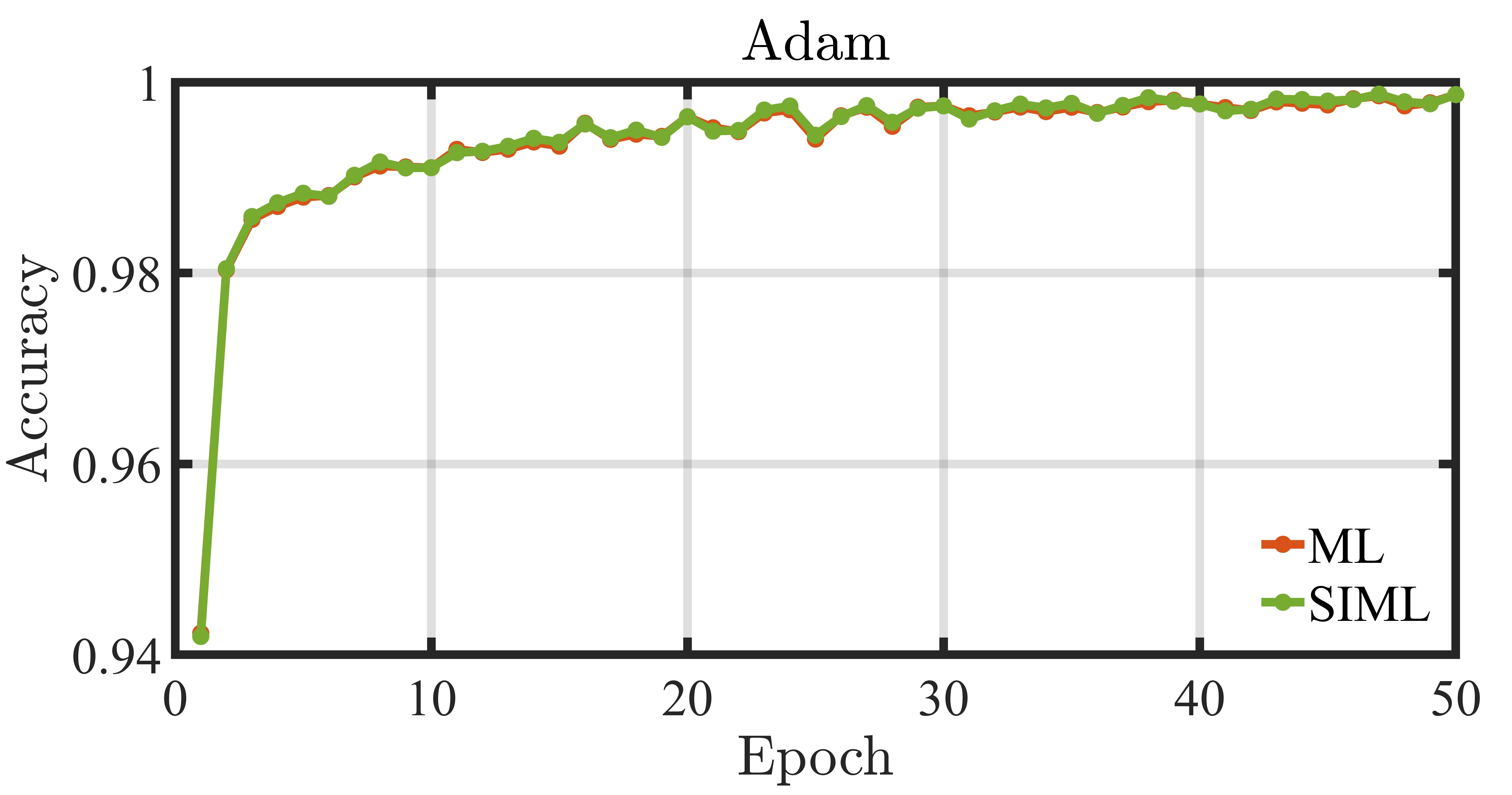}}
\caption{{The comparison of the accuracy of CNN networks in each iteration of standard ML and SIML algorithms using SGD and Adam optimizers (for ML) and target SGD and target Adam optimizers (for SIML).}}
\label{fig:accuracy}
\end{figure*}
\begin{itemize}
\item{Dataset:} We test our algorithm on the standard MNIST database for handwritten digit recognition, containing 60000 training and 10000 testing instances of 28× 28 size gray-level images, which are flattened to the vectors of dimension 784.
\item{Models:} {The MLP model is a feed-forward deep neural network with ReLU units and a softmax layer of 10 classes with two hidden layers containing 64 hidden units and cross-entropy loss containing 55,050 parameters. The CNN model has two $5 \times 5$ convolution layers (the first with 32 channels, the second with 64, each followed with $2 \times 2$ max pooling), a fully connected layer with 512 units and ReLu activation, and a final softmax output layer. The CNN model has $582,026$ weight parameters. The third model is a simple feedforward neural network built from two fully connected layers trained with $x^2$ activation function and cross-entropy loss (providing a simple implementation for homomorphic encryption operations), containing 12,730 parameters.}
\item{Optimization tools:} As optimization algorithms, the SGD and Adam optimizers with learning rate 0.001 and $T=50$ epochs are employed to train ML models.
\end{itemize}
{Our implementation uses Keras with a Tensorflow backend on an HP laptop with A100 GPU and 16 GB RAM.}\\
{Based on the dimension of flattened images in the MNIST database, the dimension of input data is $n_y=784$, which is increased to the dimension of encoded input data $\tilde{n}_y=812$ in SIML. To implement the SIML algorithm, the variables of the encoding mechanisms \eqref{distortopt1} and \eqref{distortopt2}, and target optimizer \eqref{targetoptimizer} are designed by randomly selecting small full-rank matrices $\Pi_1$-$\Pi_3$ and a large full-rank matrix $\Pi_4$ with appropriate dimensions. Then, we computed base $N_1$ of the kernel of $\Pi_{1}^{L}$. The random processes $s_i$, for $i \in \{1,2,...,n_D\}$, are defined as multivariate Laplace variables with large means and covariances. We implement various ML algorithms through standard ML, SIML, and also two well-known privacy-preserving ML tools, DP-SGD and CKKS, to compare and evaluate results. The dimension of the model parameters vector for different models and their immersed dimensions in the SIML scheme are shown in Table \ref{tab:params}.\\
\begin{table}[t]
  \caption{{Dimensions of model parameters.}}
  \centering
  \begin{tabular}{llll}
    \toprule
    \textbf{Model} & $n$ & $\Tilde{n}$\\
    \midrule
    CNN & 582,026& 592,952\\
    MLP & 55,050& 55,191\\
    NN & 12,730 & 12,758\\
    \bottomrule
  \end{tabular}
  \label{tab:params}
\end{table}
According to Theorem \ref{theoremLaplace}, given that in the SIML use case, the database $\mathcal{D}$ and the optimal trained model $w^*$ play the role of input and utility of the algorithm in immersion-based coding, the conditions:} 
\begin{equation}\label{DP-ML}{
\left\{\begin{array}{l}
\frac{||\Pi_{1}^{i}||_1 \Delta_1^{\mathcal{D}}}{ ||N_1^{i}||_2 \sigma} \le {\epsilon^{\mathcal{D}}},\,\,\,\,\,
\frac{||\Pi_{3}^{j}||_1 \Delta_1^w}{ ||\Pi_4 N_1^{j}||_2 \sigma}  \le {\epsilon^w},
 \end{array}\right.}
\end{equation}
{provide ${\epsilon^{\mathcal{D}}}$ and ${\epsilon^w}$- Differential Privacy Guarantee for each element of data $\Tilde{y}^k \in \tilde{\mathcal{D}}$, for $k \in \{1,2,...,n_D\}$ and model parameters $\Tilde{w}^*$, $\Tilde{y}^k_i$ and $\Tilde{w}^*_j$, respectively, for $i \in \{1,...,\Tilde{n}_y\}$ and $j \in \{1,...,\tilde{n}_w\}$. $\Delta_1^{\mathcal{D}}$ and $\Delta_1^w$ are the sensitivity of the database $\mathcal{D}$ and the optimal model $w^*$. Taking into account that the user database is private data, sensitivities can be calculated as $\Delta_1^{\mathcal{D}}=1$ and $\Delta_1^w=C$ \cite{abadi2016deep} where $C$ is the gradient clipping threshold to bound the gradient in gradient descent optimization algorithms. This clipping technique is used in ML with DP algorithms to be able to calculate the sensitivity of model parameters by ensuring that $\nabla l \le C$. In typical ML with DP algorithms, if the clipping threshold $C$ is too small, clipping destroys the intended gradient direction of the parameters, and if it is too large, it forces to add more noise to the parameters due to its effect on the sensitivity. But in SIML, since the distortion induced by the privacy noises can be removed by the user, these noises do not need to be small. Therefore, we can choose a very large clipping threshold to avoid distorting the ML performance. Hence, in this implementation, the clipping threshold for SIFL is $C=1000$. Considering $||\Pi_1^i||_1=10^{-4}$, $||N_1^i||_2=10^4$, $||\Pi_3^j||_1=10^{-4}$, $||\Pi_4||_2=10^4$,  $\sigma=10^4$, the ${\epsilon^{\mathcal{D}}}$ and ${\epsilon^w}$-DP guarantees for database and model parameters with ${\epsilon^{\mathcal{D}}}=1e-12$ and ${\epsilon^w}=1e-13$ can be achieved using the SIML scheme, which is a very high level of DP guarantee.}\\
\indent In Fig. \ref{fig:encodedimage}, we show the impact of the proposed coding mechanisms by comparing one sample of the original MNIST dataset and its encoded format. As can be seen in this figure, the encoded image closely resembles the random term in the distorting mechanism, differing significantly from the original image. Therefore, adversaries can not infer any information from the original image by accessing the encoded one.\\
\indent Then, the comparison between the training accuracy of the SIML and standard ML frameworks is illustrated in Fig. \ref{fig:accuracy}.  We use the CNN model (see Table \ref{tab:params}) with SGD and Adam optimizers for these implementations. Therefore, the SIML framework utilizes target SGD and target Adam optimizers \eqref{targetoptimizer} with the distorted MNIST database, while the traditional ML framework employs original SGD and Adam optimizers with the original MNIST dataset. The accuracy under the SIML configuration is similar to those in the non-privacy setting, indicating that SIML can incorporate cryptographic methods into ML systems without compromising model accuracy and convergence rate.\\
\begin{figure}[!htb]\centering
\includegraphics[width=.95\linewidth]{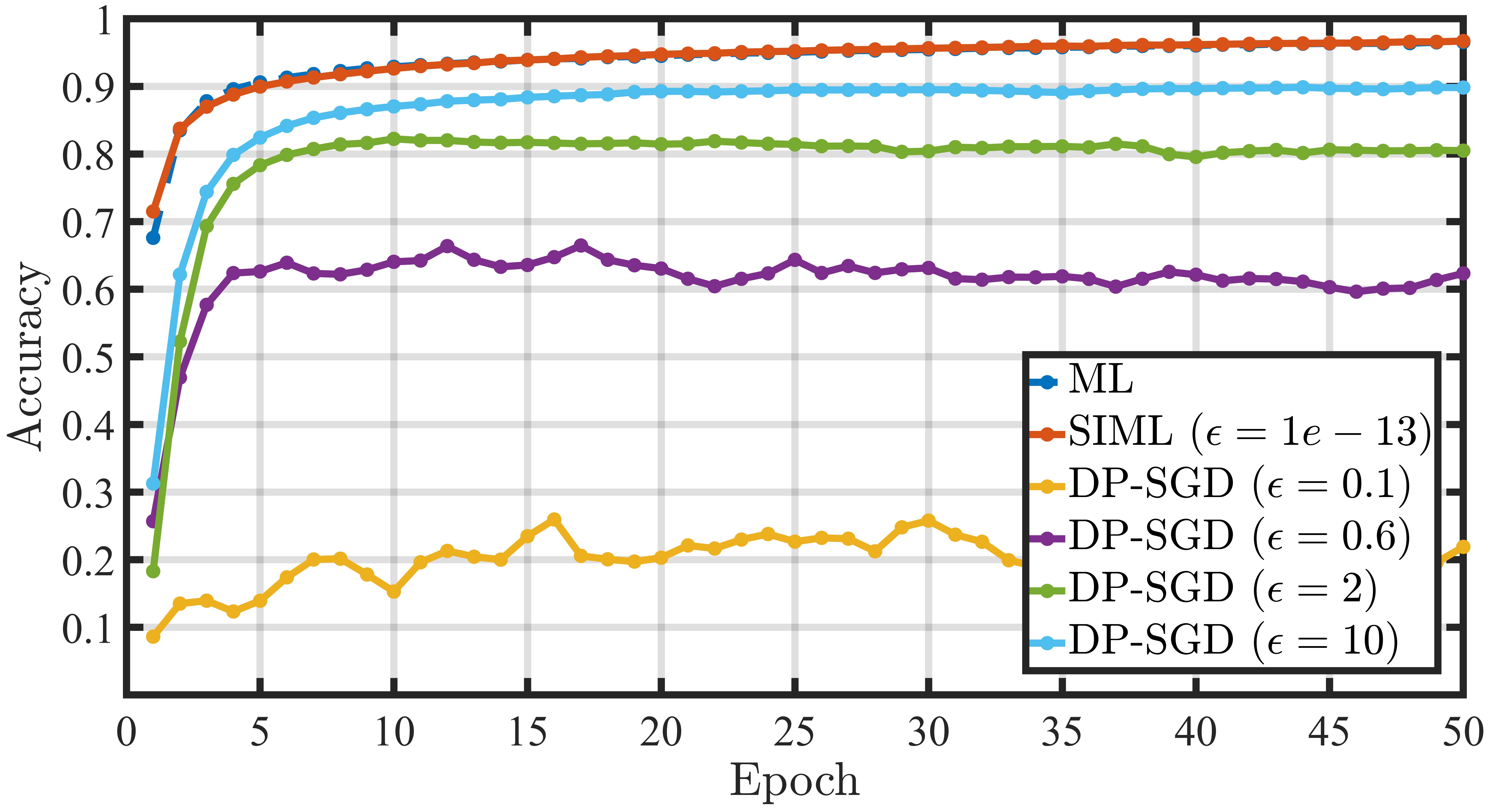}
  \caption{{The comparison of the accuracy of standard ML, SIML with privacy level ($\epsilon=1e-13$), and DP-SGD for various privacy levels $\epsilon=0.1,0.6,2,10$ and $\delta=1e-08$.}}\label{dpacccompare}
\end{figure}
\begin{figure}[!htb]\centering
\includegraphics[width=.95\linewidth]{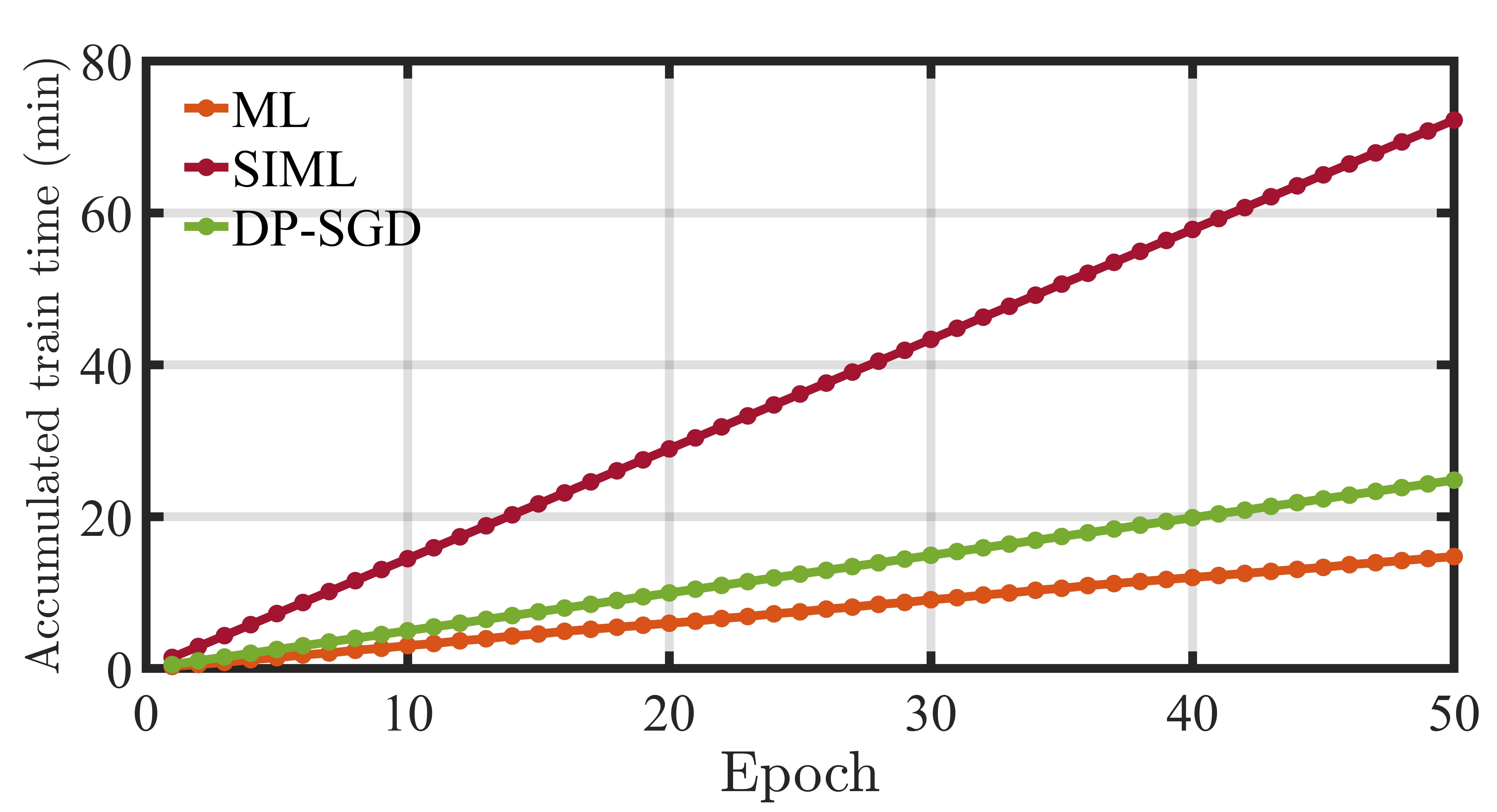}
  \caption{{The comparison of the training time of standard ML, SIML, and DP-SGD.}}\label{dptimecompare}
\end{figure}
\begin{figure}[!htb]\centering
\includegraphics[width=.95\linewidth]{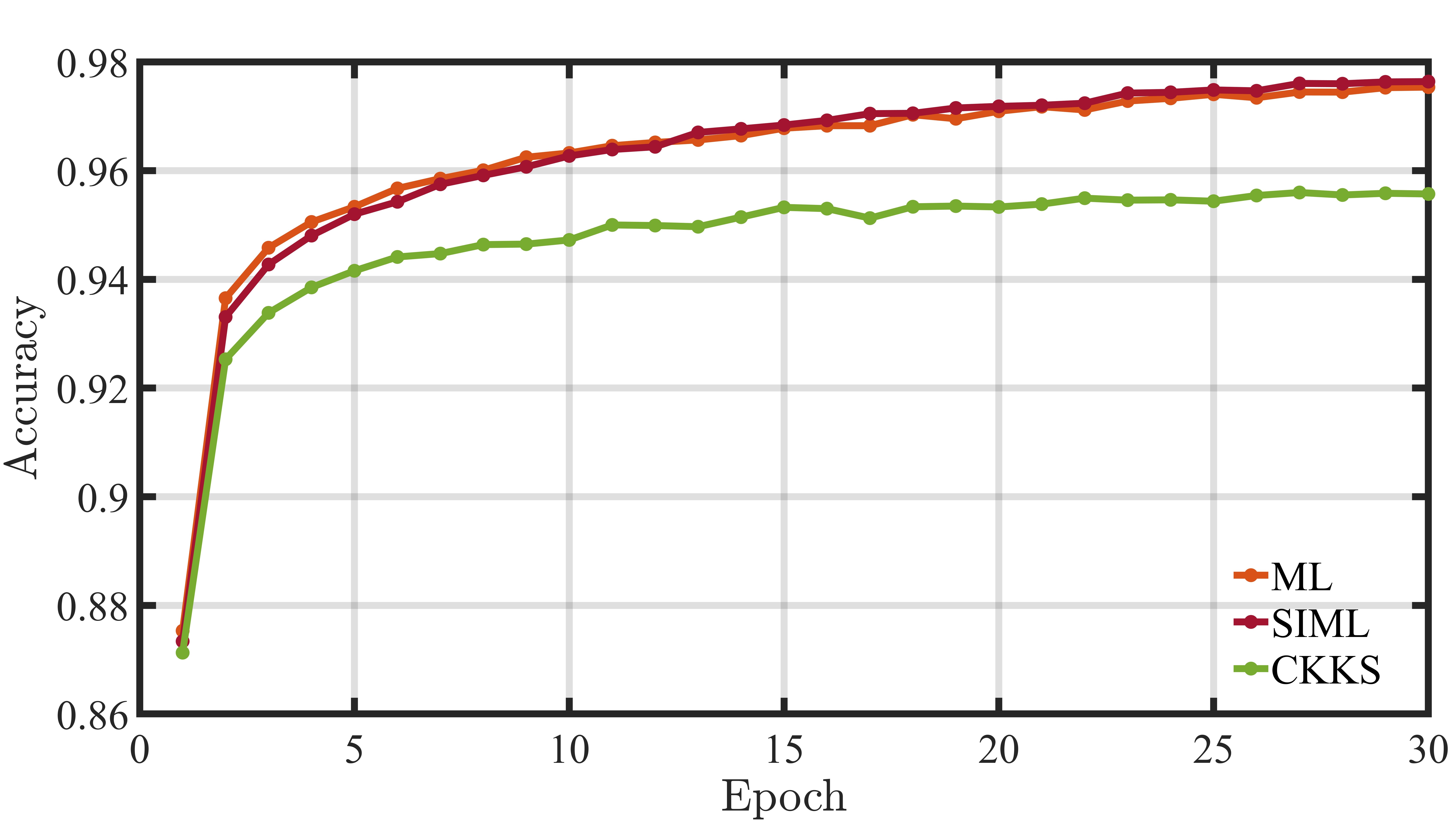}
  \caption{{The comparison of the accuracy of standard ML, SIML, and CKKS.}}\label{ckksacccompare}
\end{figure}
\begin{figure}[!htb]\centering
\includegraphics[width=.95\linewidth]{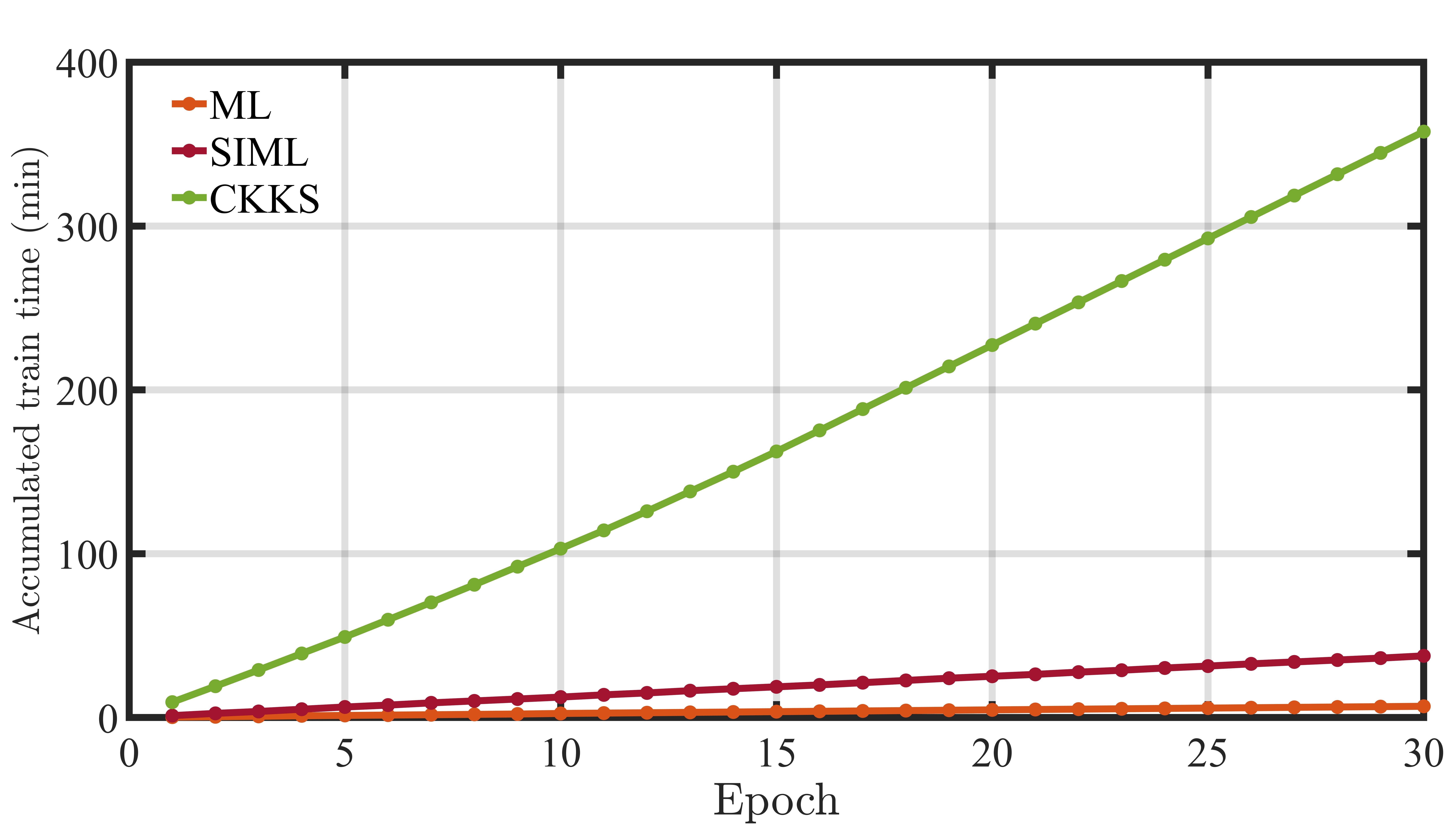}
  \caption{{The comparison of the training time of standard ML, SIML, and CKKS.}}\label{ckkstimecompare}
\end{figure}
{We evaluate the performance of the proposed SIML scheme by comparing it with two well-known privacy-preserving ML tools, DP-SGD and CKKS, in terms of computational complexity and network accuracy.\\
\indent The comparison between the accuracy and training time of SIML, a well-known differential privacy-based ML algorithm (DP-SGD), and a standard ML algorithm is illustrated in Figure \ref{dpacccompare} and \ref{dptimecompare}. For this implementation, the MLP model (see table \ref{tab:params}) is employed with normal SGD, DP-SGD, and target SGD \eqref{targetoptimizer} optimizers for training the ML, DP-SGD, and SIML models, respectively. As can be seen in Figure \ref{dpacccompare}, in the DP-SGD algorithm, a higher level of DP guarantee would lead to a significant loss in the accuracy and performance of the ML network due to distorting noises, while the proposed SIFL algorithm can provide a very high level of DP guarantee for model parameters ($\epsilon=1e-13$) without losing the accuracy and convergence rate of the model. The reason behind that is that in DP-SGD, the distortion induced by the DP noises is not removed from the model. However, in SIML, the distortion induced by the privacy noises can be removed from the final model by the user when decoding it. Therefore, a very high level of DP guarantee can be achieved without degrading the performance of the ML model. However, the comparison of the training time of ML, SIML, and DP-SGD in Figure \ref{dptimecompare} shows that the training time in SIML is higher than the training time of the DP-SGD and the standard ML algorithms which is reasonable due to the multiplication of matrices $\Pi_2$, $\Pi_2^L$, and $\Pi^L_1$ in target SGD \eqref{targetoptimizer} comparing to standard SGD \eqref{standardoptimizer}.\\
\indent In Figures \ref{ckksacccompare} and \ref{ckkstimecompare}, we compare the accuracy and training time of SIML and the CKKS encryption scheme, a form of Fully Homomorphic Encryption (FHE) designed for privacy-preserving machine learning. CKKS is particularly well-suited for machine learning tasks due to its support for approximate arithmetic on encrypted data, which is often required for operations involving real numbers. For this implementation, we use a simple NN model with the Adam optimizer, as detailed in Table \ref{tab:params}. As shown in the figures, SIML outperforms CKKS by achieving the same accuracy as the original non-private model while requiring less computational time. The accuracy loss in CKKS is due to the use of approximate arithmetic for the activation functions. Additionally, CKKS, like other FHE schemes, requires bootstrapping to refresh encoded data and manage noise. 
Therefore, the increase in computational complexity in SIML is significantly less than that associated with cryptographic tools like CKKS, because: 1)High-order polynomial approximations of nonlinear activation functions in FHE schemes lead to substantial computational overhead; and 2) FHE schemes require bootstrapping, further adding to computational costs. In contrast, immersion-based coding handles all types of linear and nonlinear activation functions directly, eliminating the need for function approximation. Additionally, bootstrapping is not required, making the immersion-based coding scheme more computationally efficient compared to other cryptographic tools. Therefore, SIML integrates cryptographic methods into machine learning systems without compromising model accuracy and maintains a reasonable computational cost.\\
\indent In summary, most current solutions for preserving the privacy of machine learning networks rely on Differential Privacy (DP) techniques, such as DP-SGD, or cryptographic tools, like CKKS. Although these methods improve user data privacy, they often result in a loss of model accuracy or increased computational and communication overhead. The proposed SIML scheme offers any desired level of DP guarantee with a fair computational cost, maintaining both model accuracy and convergence rate.}  
\subsection{Privacy-aware Networked Control}
{{The networked control system is an emerging technology that employs shared communication channels rather than specialized point-to-point connections to transmit the sensor and actuator data to close the control loop. This offers various advantages, such as easier installation and maintenance. However, control loops that outsource the computation of sensitive private data to third-party platforms via public/unsecured communication networks are already the subject of cyberattacks involving eavesdropping on private sensitive data.}
\subsubsection{Related work}
{There have been recent studies in the literature related to privacy of Networked Control Systems (NCSs) \cite{wang2024survey}. Homomorphic Encryption (HE) is a widely used method to maintain privacy in various control applications, such as encrypted linear controllers \cite{kogiso2015cyber,lin2018secure}, observer-based remote anomaly detection\cite{sadeghikhorami2021novel}, and encrypted model predictive control\cite{darup2017towards}. HE allows controllers to operate remotely without exposing sensitive and private data. However, the primary drawback of HE in encrypted control is the significant computational and communication overhead required, making it less suitable for real-time applications \cite{darup2021encrypted}. Differential Privacy (DP) is another popular method in this field, using additive random noise to enforce privacy \cite{koufogiannis2017differential}. DP-based mechanisms have been proposed for applications in consensus, filtering, distributed optimization, and LQ control \cite{cortes2016differential, yazdani2022differentially}. However, added noise in the DP can degrade the accuracy of the control system, potentially affecting its performance and decision-making capabilities.\\
\indent In summary, existing privacy-preserving methods for NCSs often result in performance loss or increased computational overhead. In this section, we apply the immersion-based coding proposed in Proposition 1 to enable secure networked control. Results demonstrate that the immersion-based coding scheme can preserve the privacy of NCSs without degrading controller performance and with minimal computational cost, and it can be applied to complicated nonlinear NCSs.}
\subsubsection{Case Study of Networked Control Systems}
We illustrate the performance of System Immersion-based coding for Networked Control Systems (SI-NCS) through a case study of a two-stage chemical reactor with delayed recycle streams. The authors in \cite{jia2017adaptive} propose an adaptive output feedback controller for this chemical reactor dynamics. We assume the controller is run in the cloud and aim to run it in a private manner using the proposed coding scheme. Consider the dynamic model of the reactor (as reported in \cite{jia2017adaptive}):
\begin{figure*}[t]
  \centering
  \subcaptionbox{}[.45\linewidth][c]{%
    \includegraphics[width=1\linewidth]{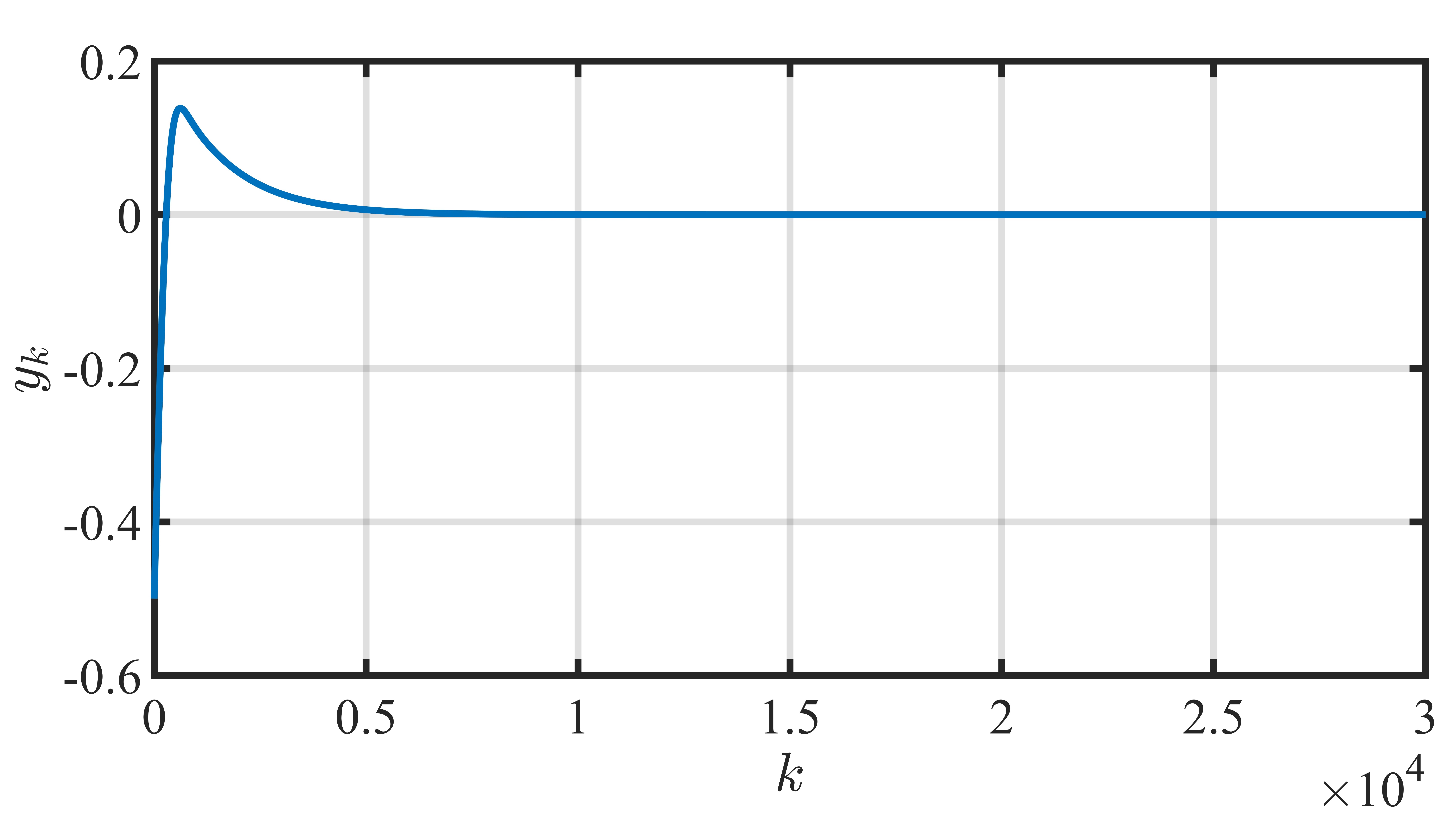}}\quad
  \subcaptionbox{}[.46\linewidth][c]{%
    \includegraphics[width=1\linewidth]{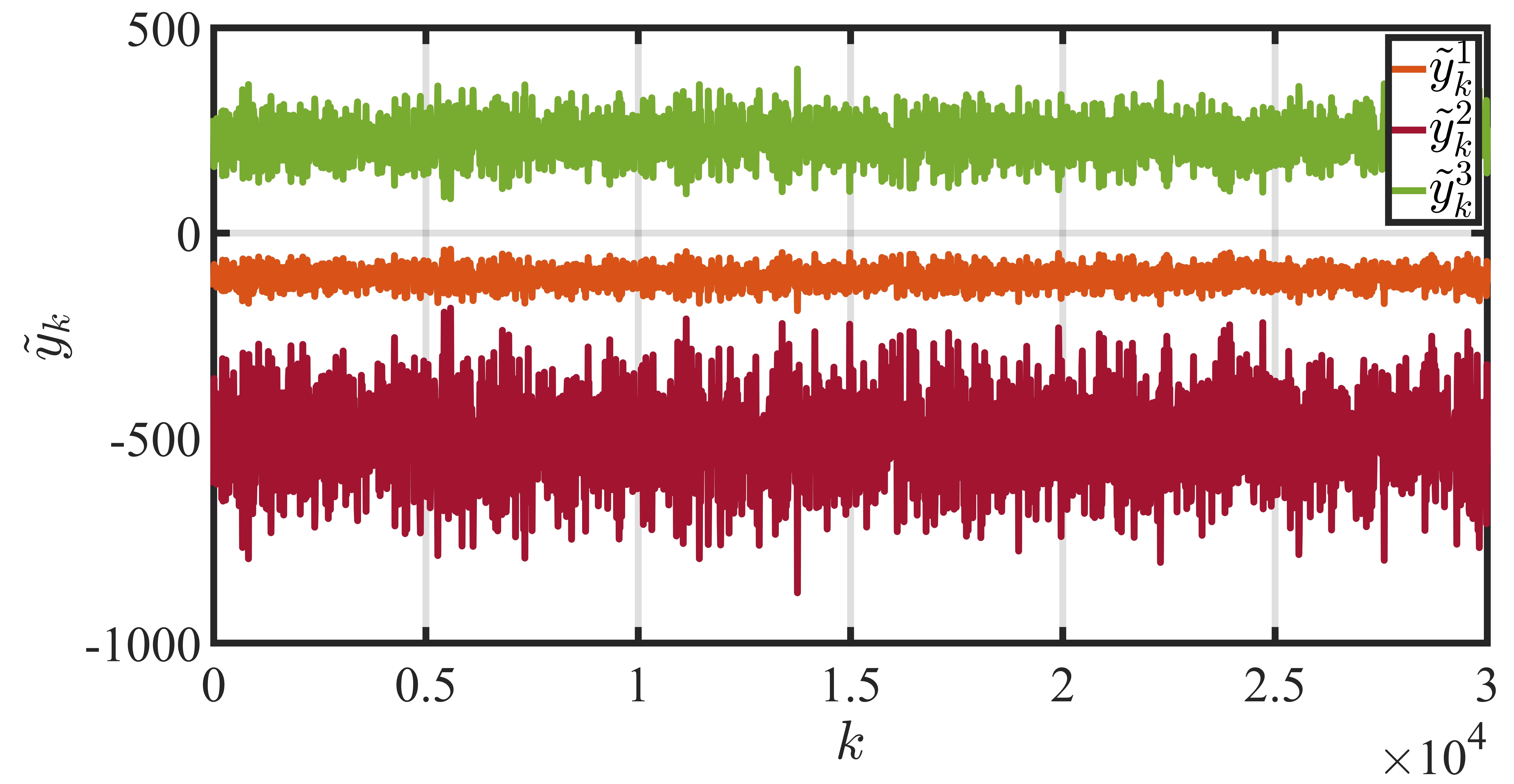}}
    \caption{Comparison between the measurement vector $y_k$ and the encoded measurements $\tilde{y}_k$.}
\label{yyp}
\end{figure*}
\begin{equation}\label{systemofcontroller}
\left\{\begin{array}{l}\begin{aligned}
{x}^{1}_{k+1}&=-(\frac{1}{\theta_{1}}+k_1) x^{1}_k+\frac{1-R_{2}}{V_{1}} x^{2}_k+\delta_{1}(\vartheta, x_{\tau}), \\
{x}^{2}_{k+1}&=-(\frac{1}{\theta_{2}}+k_2) x^{1}_k+\frac{R_{1}}{V_{2}} x^{1}_{\tau}+\frac{F_{2}}{V_{2}} u_k+\delta_{2}(\vartheta, x_{\tau}), \\
y_k&= x^{1}_k,\end{aligned}
\end{array}\right.
\end{equation}
with system state $x_k=[x^1_k,x^2_k]$, controller $u_k$, and measurement $y_k$, delayed time index $\tau=k-d_k$ with unknown time-varying delay $d_k$, parameters $\theta_{1}=\theta_{2}=2$, $k_{1}=k_{2}=0.3$, $R_{1}=R_{2}=0.5$, $V_{1}=V_{2}=0.5$, $F_{2}=0.5$, and nonlinear functions with unknown parameters $\vartheta_{i}$, $i=1,2,3$, as follows:
\begin{equation}\label{unknownparams}
\left\{\begin{array}{l}\begin{aligned}
\delta_{1}&=\vartheta_{1} \sin (k) (x^{1}_{\tau})^{2},\\
\delta_{2}&=\vartheta_{2} \sin (k) (x^{1}_{\tau})^{3}+\vartheta_{3} x^2_k.
\end{aligned}
\end{array}\right.
\end{equation}
We use the following adaptive output feedback controller (see \cite{jia2017adaptive} for details) in order to achieve global adaptive stabilization of system \eqref{systemofcontroller}:
\begin{equation} \label{dynamicalcontroller}
\left\{\begin{array}{l}\begin{aligned}
{\hat{z}}_{k+1}&=-r_k\left(\hat{z}_k+r_k y_k+l_k \rho_k y_k\right)-r_k {\hat{z}}_k-r_k^{2} y_k\\&-\max \{r_k \Delta_k-r_k^2, \rho_k y_k^2+\zeta_k^2\} y_k, \\
{r}_{k+1}&=\max \{r_k \Delta_k-r_k^2, \rho_k y_k^2+\zeta_k^2\}, \\
{l}_{k+1}&=\rho_k y_k^{2}, \\
u_k&=-r_k\left(\hat{z}_k+r_k y_k+l_k \rho_k y_k\right),
\end{aligned}
\end{array}\right.
\end{equation}
where $\hat{z}_0=0$, $r_0=l_0=1$, and
\begin{equation}
\left\{\begin{array}{l}
\begin{aligned}\label{eq41}
\rho_k&=\left(0.1+y_k^2\right)^{2}+1.01, \\
\zeta_k&=r_k^{-\frac{1}{2}}\left(\hat{z}_k+r_k y_k+l_k \rho_k y_k\right),\\
\Delta_k&=2\left(2 l_k^{2}+ l_k^{4}\left(1.02+0.2 y_k^{2}+ y_k^{4}\right)\right) (1.0404\\&+1.224 y_k^{2}+10.56 y_k^{4}+6 y_k^{6}+25 y_k^{8})\\&+4 y_k^{4}\left(1.02+0.2 y_k^{2}+y_k^{4}\right)^{4}.
\end{aligned}
\end{array}\right.
\end{equation}

We consider the setting where the controller \eqref{dynamicalcontroller}, \eqref{eq41} is run in the cloud. We employ the immersion-based coding proposed in Proposition \ref{proposition1} to enable a secure networked control system. The controller \eqref{dynamicalcontroller} can be reformulated in a compact form as follows.
\begin{equation}
\mathcal{C}:\left\{\begin{array}{l}\begin{aligned}
{x}^c_{k+1}&=f_c\left(x^c_k,y_k\right), \\
u_k&=g_c\left(x^{c}_k, y_k\right),\end{aligned}
\end{array}\right.
\end{equation}
where $x^c_k:=[\hat{z}_k,r_k,l_k]^T$ denotes the controller state vector, and $f_c(\cdot)$ and $g_c(\cdot)$ are nonlinear functions that can be defined based on \eqref{dynamicalcontroller}. Therefore, controller $\mathcal{C}$ can be considered as dynamical system $\Sigma$ in \eqref{generaldynamics} that we want to immerse to provide privacy. Consequently, $x^c_k$, $y_k$, and $u_k$ are equivalent to system internal variables $\zeta_k$, input $y_k$, and utility $u_k$ in system $\Sigma$. Hence, using the solution in Proposition \ref{proposition1}, the encoding maps for the controller $u_k$ and measurement $y_k$, the immersion map of the controller state $x^c_k$, and the target dynamical controller $\tilde{\mathcal{C}}$ can be designed as follows.\\
The encoding and immersion maps are given by:
\begin{equation}\label{privacymechanismscontrol}
\left\{\begin{array}{l}\begin{aligned}
\tilde{y}_k&=\Pi_1 y_k + N_1 s_k,\\
\tilde{x}^c_k&=\Pi_2 x^c_k,\\
\tilde{u}_k &= \Pi_3 u_k + \Pi_4 \tilde{y}_k,
\end{aligned}
\end{array}\right.
\end{equation}
the target controller is given by:
\begin{equation}\label{targetalgorithmsolutioncontrol}
\tilde{\mathcal{C}}=\left\{\begin{array}{l}\begin{aligned}
{\tilde{x}}^c_{k+1}&=\tilde{f}_c\left({\tilde{x}}^c_k, \tilde{y}_k\right)=\Pi_{2} f_c\left(\Pi_2^L {\tilde{x}}^c_k,\Pi_1^L \tilde{y}_k\right), \\
\tilde{u}_k&=\tilde{g}_c\left({\tilde{x}}^c_k, \tilde{y}_k\right) \\
&= \Pi_3 {g}_c\left(\Pi_2^L {\tilde{x}}^c_k,\Pi_1^L \tilde{y}_k\right)+ \Pi_4 \tilde{y}_k,\end{aligned}
\end{array}\right.
\end{equation}
and the inverse function:
\begin{equation}\label{inversesolutioncontrol}
\pi_3^L\left(\tilde{u}_k,\tilde{y}_k\right) := \Pi_3^L \left( \tilde{u}_k-\Pi_4 \tilde{y}_k\right),
\end{equation}
with \emph{full rank matrices} $\Pi_1 \in \mathbb{R}^{\tilde{n}_y \times n_y}$, $\Pi_2 \in \mathbb{R}^{\tilde{n}_{x_c} \times n_{x_c}}$, $\Pi_3 \in \mathbb{R}^{\tilde{n}_u \times n_u}$, $\Pi_4 \in \mathbb{R}^{\tilde{n}_u \times \tilde{n}_y}$, matrix $N_1 \in \mathbb{R}^{\tilde{n}_y \times (\tilde{n}_y - n_y)}$ expanding the kernel of $\Pi_{1}^{L}$, and a random vector $s^1(t) \in \mathbb{R}^{(\tilde{n}_y - n_y)}$.\\
{Our implementation uses MATLAB version R2023a Simulink on an HP laptop with 16 GB RAM.}\\
The dimensions of vectors in the original controller \eqref{dynamicalcontroller} are $n_{x_c}=3$ and $n_{y}=n_u=1$. For the design of the target controller, we consider the immersion dimensions $\tilde{n}_{x_c}=4$ and $\tilde{n}_y=\tilde{n}_u=3$. Also, for the design of mappings' matrices, we randomly select small full-rank matrices $\Pi_1$-$\Pi_3$ and a large full-rank matrix $\Pi_4$ based on the designed dimensions. Then, we compute the base $N_1$ of the kernel of $\Pi_{1}^{L}$. The random process $s_k$ is defined as a multivariate Laplace variable with a large mean and covariance. {According to Theorem \ref{theoremLaplace}, given that in the SI-NSC use case, the measurement $y_k$ and the control action $u_k$ play the role of input and utility of the algorithm in immersion-based coding, the conditions
$\frac{||\Pi_{1}^{i}||_1 \Delta_1^{y}}{ ||N_1^{i}||_2 \sigma} \le {\epsilon^{y}}$ and $\frac{||\Pi_{3}^{j}||_1 \Delta_1^u}{ ||\Pi_4 N_1^{j}||_2 \sigma}  \le {\epsilon^u}$, provide $\epsilon^y$ and ${\epsilon^u}$- Differential Privacy Guarantee for each element of the encoded measurement $\Tilde{y}_k$ and control action $\Tilde{u}_k$, $\Tilde{y}^i_k$ and $\Tilde{u}^j_k$, respectively, for $i \in \{1,...,\Tilde{n}_y\}$ and $j \in \{1,...,\tilde{n}_u\}$. $\Delta_1^y$ and $\Delta_1^u$ are the sensitivity of the measurement data and the control action. Considering $y_k$ and $u_k$ as private data of the NCS, the sensitivities are equivalent to $\Delta_1^y=1$ and $\Delta_1^u=1$. To design additive privacy noises to achieve DP guarantee, since in SI-NCS, the distortion induced by the privacy noises can be removed by the user, these noises do not need to be small. Hence, considering $||\Pi_1^j||_1=10^{-4}$, $||N_1^j||_2=10^4$, $||\Pi_3^j||_1=10^{-4}$, $||\Pi_4||_2=10^4$,  $\sigma=10^4$, the $\epsilon^y$ and ${\epsilon^u}$-DP guarantees for measurement data and control action with $\epsilon^y=1e-12$ and ${\epsilon^w}=1e-16$ can be achieved by SI-NCS scheme, which is a very high level of DP-guarantee.}\\
\indent For implementing the dynamic model of the reactor given in \eqref{systemofcontroller} and \eqref{unknownparams}, we consider time-varying delay $d_k=0.5(3+\sin (k))$, initial states $x^1_0=-0.5$, $x^2_0=2$, and unknown parameters $\vartheta_1=\vartheta_2=1$, $\vartheta_3=-1$.\\
\indent First, in Fig. \ref{yyp}, we show the effect of the proposed coding mechanisms, where we contrast actual and encoded measurement data. As illustrated in this figure, when $\Pi_1$ is very small and $s_k$ is large according to \eqref{privacymechanismscontrol}, the encoded measurement $\tilde{y}_k$ closely resembles the random component, differing significantly from the actual measurement $y_k$. Moreover, the dimension of the measurement vector $y_k$ expands from one to three in the encoded form $\tilde{y}_k$. Therefore, adversaries can not even access the true dimension of the measurement vector. A similar outcome is observed when comparing the actual and encoded control actions $u_k$ and $\tilde{u}_k$.\\
\begin{figure}[!htb]
\centering
\includegraphics[width=.99\linewidth]{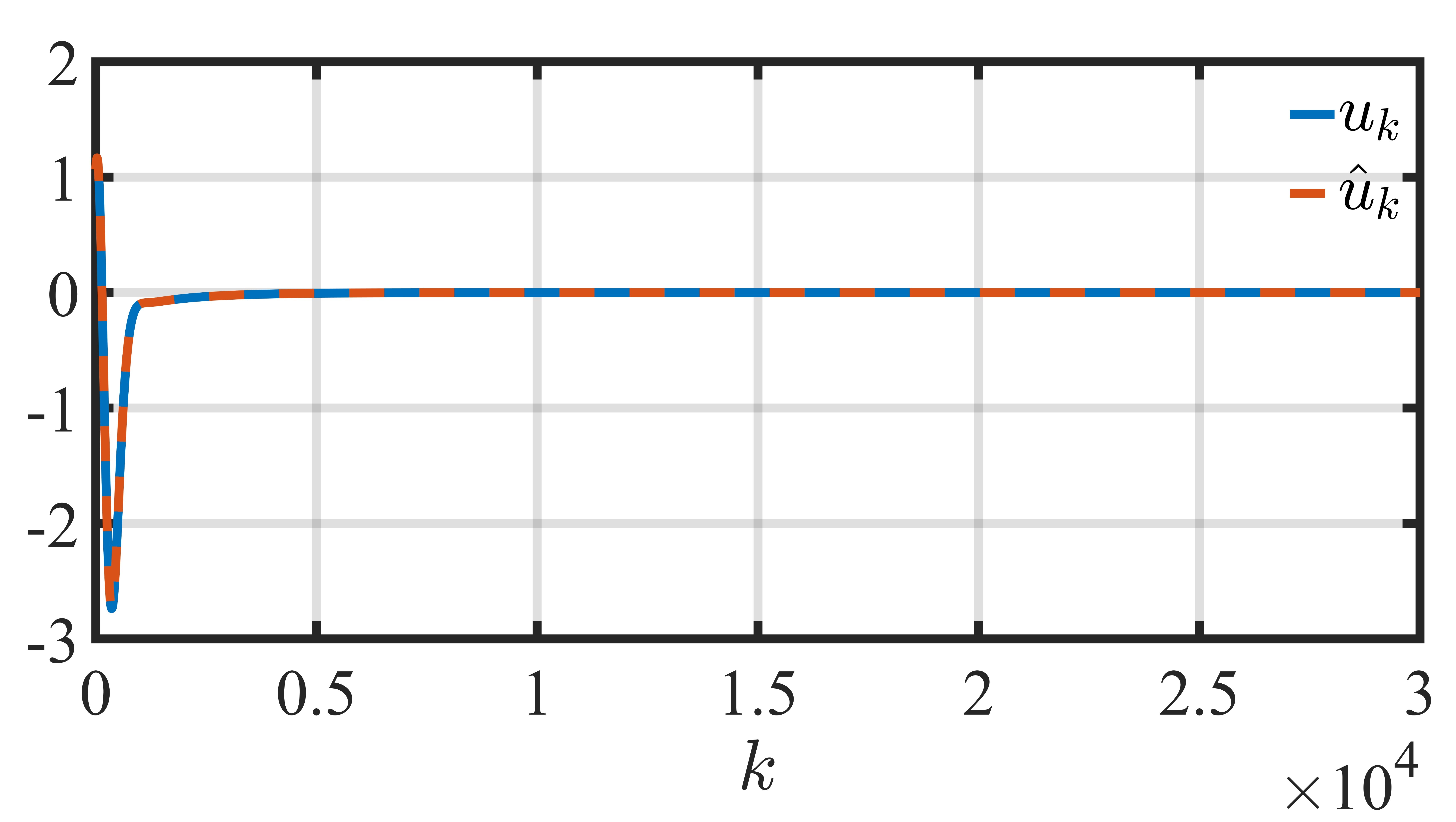}
  \caption{Comparison between the control action $u_k$ and the decoded control action $\hat{u}_k$.}\label{uup}
\end{figure}
\indent In Fig. \ref{uup}, the original control action $u_k$ and the decoded control action at the user side $\hat{u}_k$ are compared. As can be seen, the decoded control action is almost identical to the original control action, and the error (due to matrix products) between the original and decoded control action, which is shown in Fig. \ref{erroruup}, is negligible. Hence, we can conclude that the proposed immersion-based coding can integrate a cryptographic method in the networked control systems without sacrificing the performance of the controller, and there is no need to hold a trade-off between privacy and control performance.\\
\indent Therefore, although encoded private signals $\tilde{y}_k$ and $\tilde{u}_k$ are different from original signals $y_k$ and $u_k$, the decoded controller at the user's side, $\hat{u}_k$, is same as $u_k$. Consequently, the proposed privacy-preserving networked control system shows equivalent control performance to the standard controller.\\
\begin{figure}[!htb]\centering
\includegraphics[width=.99\linewidth]{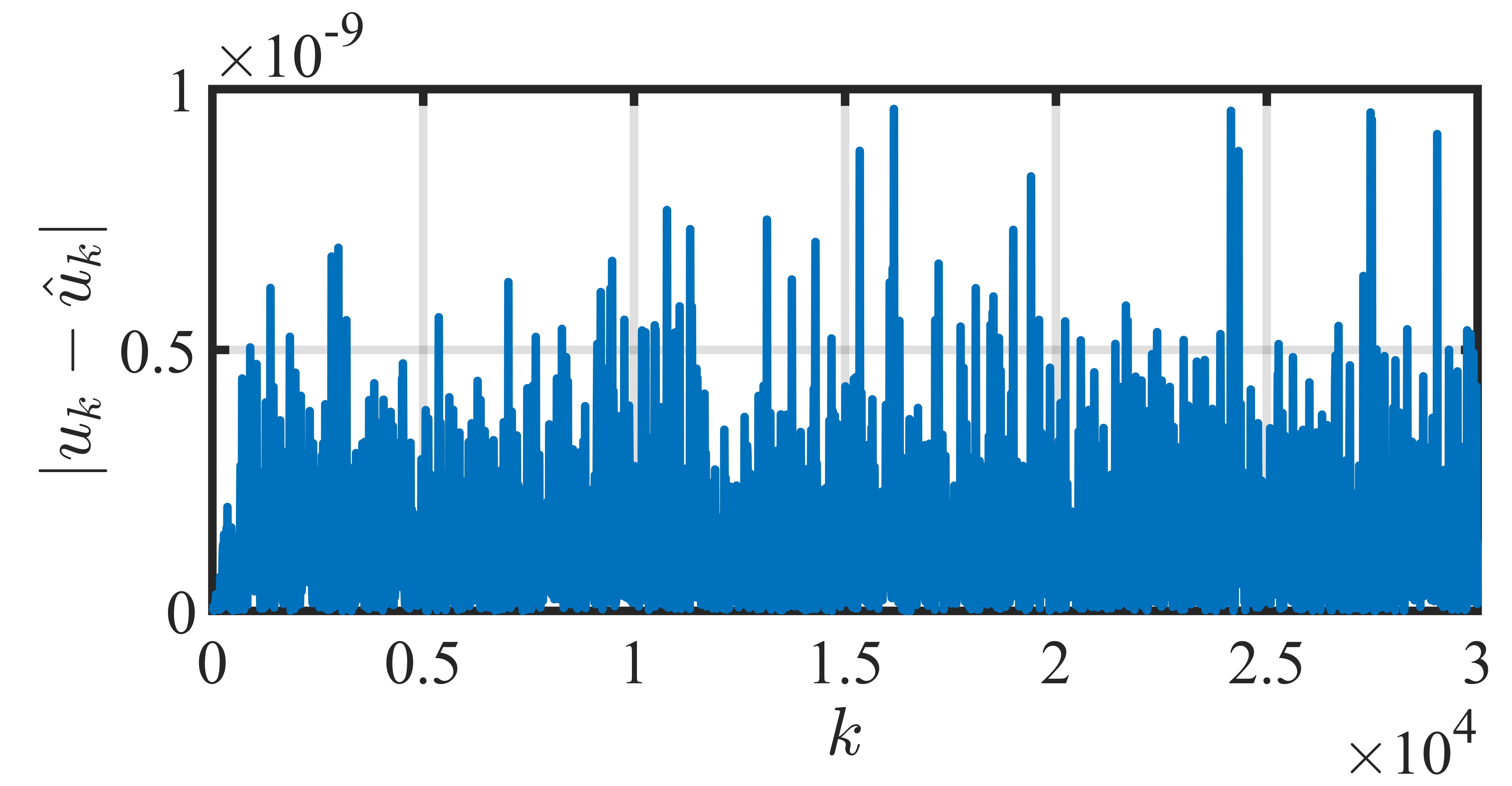}
  \caption{The error between the control action $u_k$ and the decoded control action $\hat{u}_k$.}\label{erroruup}
\end{figure}
\indent Fig. \ref{xxh} compares the states of the closed-loop system using the original control action $u_k$, $x_k^1$ and $x_k^2$, and the states of the closed-loop system using the decoded control action $\hat{u}_k$, $\hat{x}_k^1$ and $\hat{x}_k^2$. It can be clearly observed from Fig. \ref{xxh} that the states of the system with and without the immersion-based coding are almost identical, and asymptotic stabilization has been achieved in both cases with good robustness against the time-varying delay.\\
\indent {Finally, the comparison between the accumulated running time of the non-private Networked Control System (NCS) and the private System Immersion-based Networked Control system (SI-NCS) is illustrated in Figure \ref{controltraintime}. As can be seen, the increased running time of the SI-NCS
in the cloud compared to the running time of the original non-private NCS is negligible.}
\begin{figure}[!htb]\centering
\includegraphics[width=.99\linewidth]{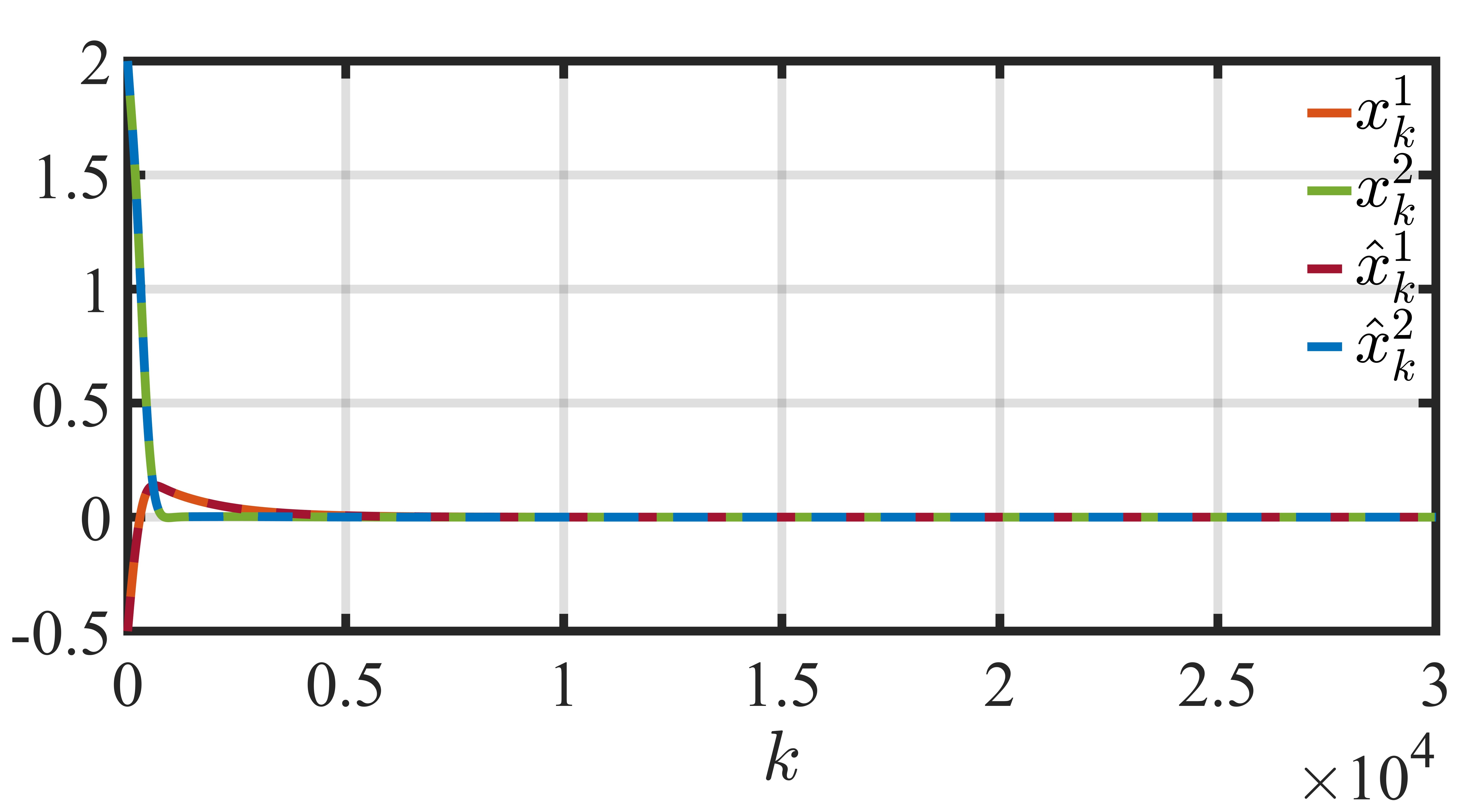}
  \caption{Comparison between closed-loop system states using the original control action $u_k$, $x_k^1$ and $x_k^2$, and closed-loop system states using the decoded control action $\hat{u}_k$, $\hat{x}_k^1$ and $\hat{x}_k^2$.}\label{xxh}
\end{figure}
\begin{figure}[!htb]\centering
\includegraphics[width=.99\linewidth]{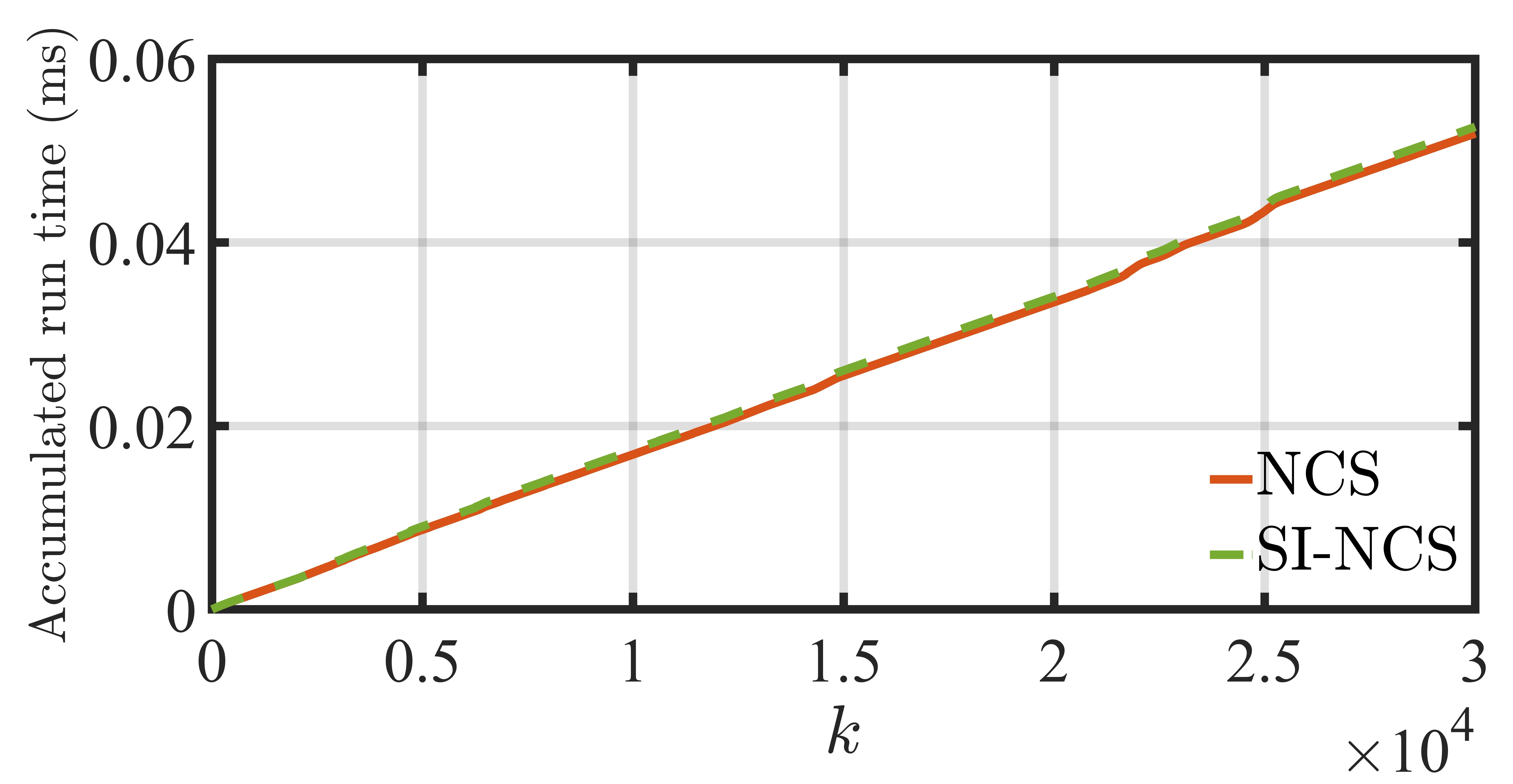}
  \caption{{Comparison between running time of networked control system without privacy (NCS) and with immersion-based privacy (SI-NCS).}}\label{controltraintime}
\end{figure}
\section{Conclusion and Future Works}\label{sec7}
In this paper, we have developed a privacy-preserving framework for the implementation of remote dynamical algorithms in the cloud. It is built on the synergy of random coding and system immersion tools from control theory to protect private information. We have devised a synthesis procedure to design the dynamics of a coding scheme for privacy and a higher-dimensional system called target algorithm such that trajectories of the standard dynamical algorithm are immersed/embedded in its trajectories, and it operates on randomly encoded higher-dimensional data. Random coding was formulated at the user side as a random change of coordinates that maps original private data to a higher-dimensional space. Such coding enforces that the target algorithm produces an encoded higher-dimensional version of the utility of the original algorithm that can be decoded on the user side.\\
\indent The proposed immersion-based coding scheme provides the same utility as the original algorithm (i.e., when no coding is employed to protect against data inference), (practically) reveals no information about private data, can be applied to large-scale algorithms, is computationally efficient, and offers any desired level of differential privacy without degrading the algorithm utility.\\ 
\indent {Regarding future works, this study has utilized affine maps to design encoding schemes, but it lacks a method for constructing general nonlinear encoding maps. Future research could focus on developing design methods for nonlinear maps using control theory for model-based approaches and machine learning tools for data-based approaches. Additionally, the current approach guarantees differential privacy on an element-wise basis for user data and utility. Future designs should aim to ensure differential privacy over the entire vector of user and utility data. Furthermore, the current immersion-based coding scheme encounters significant computational complexity with very large algorithms involving millions of parameters. Future investigations should explore optimizing the structure of encoding matrices to enhance efficiency for such problem sizes.} 
\bibliographystyle{IEEEtran}
\bibliography{ifacconf32}

\vfill

\end{document}